# Understanding the onset of hot streaks across artistic, cultural, and scientific careers

Lu Liu[1,2,3,4], Nima Dehmamy[1,2,3], Jillian Chown[1,3], C. Lee Giles[4,5], and Dashun Wang[1,2,3,6*]

[1]Center for Science of Science and Innovation, Northwestern University, Evanston, IL, USA.

[2]Northwestern Institute on Complex Systems, Northwestern University, Evanston, IL, USA.

[3]Kellogg School of Management, Northwestern University, Evanston, IL, USA.

[4]College of Information Sciences and Technology, Pennsylvania State University, University Park, PA, USA

[5]Department of Computer Science and Engineering, Pennsylvania State University, University Park, PA, USA

[6]McCormick School of Engineering, Northwestern University, Evanston, IL, USA.

**Abstract**

Hot streaks dominate the main impact of creative careers. Despite their ubiquitous nature across a wide range of creative domains, it remains unclear if there is any regularity underlying the beginning of hot streaks. Here, we develop computational methods using deep learning and network science and apply them to novel, large-scale datasets tracing the career outputs of artists, film directors, and scientists, allowing us to build high-dimensional representations of the artworks, films, and scientific publications they produce. By examining individuals' career trajectories within the underlying creative space, we find that across all three domains, individuals tend to explore diverse styles or topics before their hot streak, but become notably more focused in what they work on after the hot streak begins. Crucially, we find that hot streaks are associated with neither exploration nor exploitation behavior in isolation, but a particular sequence of exploration followed by exploitation, where the transition from exploration to exploitation closely traces the onset of a hot streak. Overall, these results unveil among the first identifiable regularity underlying the onset of hot streaks, which appears universal across diverse creative domains, suggesting that a sequential view of creative strategies that balances experimentation and implementation may be particularly powerful for producing long-lasting contributions, which may have broad implications for identifying and nurturing creative talents.

[*] To whom correspondence should be addressed. E-mail: dashun.wang@northwestern.edu

A remarkable feature of creative careers is the existence of hot streaks, which are bursts of high-impact works clustered together in close succession[1-3]. Despite the ubiquitous nature of hot streaks across artistic, cultural, and scientific domains, it remains unclear if there are any regularities underlying the beginning of a hot streak. Understanding the origin of hot streaks is not only crucial for our quantitative understanding of patterns governing creative life cycles; it also has implications for the identification and development of talent across a wide range of settings[4,5]. Deciphering what predicts hot streaks, however, remains a challenge, partly due to the complex nature of creative careers[1,6-16]. The lack of systematic explanations for hot streaks, combined with the randomness of when they occur within a career[1], paints an unpredictable, if incomplete view of creativity across a diverse range of domains.

Of the myriad forces that might affect career progression and success, the strategies of exploration and exploitation have attracted enduring interests from a broad set of disciplines[14-21], prompting us to examine their potential relationship with hot streaks. Indeed, according to the literature, exploitation allows individuals to build knowledge in a particular area and to refine their capabilities in that area over time. This could be relevant for understanding hot streaks since exploitation allows individuals to "go deep" in a focal area to both establish expertise in that area and foster a reputation related to that expertise[17,18]. Exploration, on the other hand, engages individuals in experimentation and search beyond their existing or prior areas of competency. Although exploration is more risky and consequently associated with larger variance in outcomes[22], it may also increase one's likelihood of stumbling upon a groundbreaking idea through unanticipated combinations of disparate sources[23]. In contrast, exploitation, as a conservative strategy, may stifle originality and, may over time, limit an individual's ability to consistently produce high-impact work[14]. Taken together, the benefits and downsides to these contrasting approaches raise a fundamental question: Are career hot streaks reflective of exploration or exploitation behavior, or some combination of the two?

To answer this question, we develop computational methods using deep learning[24,25] and network science[26,27] and apply them to novel, large-scale datasets tracing the career outputs of artists, film directors, and scientists. Specifically, we build high-dimensional representations of the artworks, films, and scientific publications they produce (Supplementary Information (SI) S1), which capture abstract concepts, styles, and topics represented therein, allowing us to trace an individual's career trajectory on the underlying creative space (SI S1). We further quantify the hot streak within each career by the impact of works one produced[1], measured by auction price[1,28], IMDB ratings[1,29], and paper citations in ten years[1,12], respectively. We then correlate the timing of hot streaks with the creative trajectories for each individual, allowing us to examine changes in the characteristics of the work one produces around the beginning of a hot streak.

To examine the art styles of each artist and their exploration and exploitation dynamics, we collected over 800K images of visual arts from museum and gallery collections, covering career histories of 2,128 artists[30,31]. Building on recent advances on computer vision[32,33], we use a transfer-learning approach[34] to construct an embedding for artworks using deep neural network. Specifically, we apply a pre-trained VGGNet algorithm[32], one of the best known algorithms for image recognition, to images of artworks, and connect it with an additional neural network with fully-connected layers to classify the art style labels recorded in our dataset (Fig. 1a). The convolutional layers in the pre-trained model use 3x3 filters to detect local patterns from the artwork (Fig. 1b). The filters in the first layer capture spatial patterns such as line orientations and brush strokes (Fig. 1b), whereas those in higher layers combine outputs of filters from lower layers to capture more complex features, such as shapes and objects (Fig. 1c). To leverage VGGNet's image recognition capabilities, here we do not train the VGGNet layers, but instead train the fully-connected layers to repurpose VGGNet to identify art styles (Fig. 1a), helping the first two fully-connected layers to find an abstract representation of concepts and themes by grouping together related outputs of the VGGNet layers.

Prior research shows that art style may be decoded from both brush strokes and the overall concepts, subjects and themes[33,35], suggesting that both low- and high-level features are important for capturing art styles. To this end, we combine the outputs from the first and third convolutional layers in VGGNet with the fully-connected layer before the final classification layer, and use principal component analysis (PCA) for dimensionality reduction to generate a 200-dimensional embedding of each artwork (see SI S1.1 for several case studies showing how art styles are interpreted by our deep learning framework). We then apply our deep neural network to career outputs of each artist in the dataset, and identify art styles through clusters on the 200-dimensional embedding space, allowing us to trace the evolution of art styles over the course of their careers (Fig. 2a-d).

To examine career histories of film directors, we collected our second dataset capturing plot description and cast information for each film recorded in the IMDB database (79k films by 4,337 directors; see SI S4 for more detail). From this, we build a high-dimensional representation of a film by combining its plot and cast information. We first train word embeddings[36] in the description of the plot to learn a 100-dimensional text representation of a film from co-occurrence of words (Fig. 1d) (SI S1.2). To incorporate casting information, we construct a weighted co-casting network among all actors, and apply a node embedding method DeepWalk[37] to obtain a 100-dimensional casting vector for each film (Fig. 1e) (SI S1.2). We then concatenate the vectors for plot and cast, allowing us to develop a 200-dimensional embedding space to represent all films. Despite the myriad factors that may affect the artistic and financial success of a film[38], ranging from the screenplay to acting, we find that the learned high-dimensional representation can successfully predict film genre with an accuracy of 0.948 (SI S1.2). Finally, we classify each film based on clusters in the obtained embedding space, allowing us to investigate the dynamics of styles for film directors (Fig. 2e-h).

In the third setting, we analyze career histories of 20,040 scientists by combining publication and citation datasets from the Web of Science (WoS) and Google Scholar (GS)[1,12], tracing the dynamics of research topics as reflected in the publication history of each career. We use a method developed recently by Zeng and colleagues[16] which identifies research topics within a career by finding communities in a weighted co-citing network of all publications by the individual (Fig. 1f, Fig. 2i-l). To ensure that the results obtained for scientific careers are consistent with the embedding methods used to analyze the careers of artists and directors, we also applied a node embedding method to the co-citing network to identify research topics, and repeated all our analyses, finding our conclusions remain the same (SI S1.3).

To quantify the exploration and exploitation behaviors reflected in each individual's career across the three domains, we measure the style or topic entropy for the work one produces, defined as $\widetilde{H} = -\sum_{i=1}^{m} p_i \log p_i$, where $p_i$ is the frequency in which one devotes to an art style or topic $i$ and $m$ is the number of unique styles or topics. On one extreme, a pure exploitation strategy means that an individual's work is contained within only one topic or style ($\widetilde{H} = 0$); on the other extreme, $\widetilde{H} = \log n$ corresponds to the case of pure exploration, where $n$ is the number of works one produced in the period, indicating that an individual's attention is evenly divided across a distribution of topics ($p_i = 1/n$). For convenience, we normalize the entropy measure to obtain the rescaled entropy $H = \widetilde{H}/\log n$. Fig. 2 illustrates three notable careers as examples for identifying art styles, topics and their entropies calculated using the methodologies described above.

To test whether hot streaks are associated with exploration or exploitation, we measure the distribution of entropy $P(H)$ for works produced before and during a hot streak (Fig. 3a-c). To gauge the expected magnitude of $H$ around a hot streak, we further construct a null model for each career by randomly designating the time at which the hot streak begins[1]. We calculate the average entropy $\langle H \rangle$ measured in real careers before (Fig. 3d-f) and after the onset of hot streak (Fig. 3g-i), and compare them with random careers, measured by the distribution of entropy, $P(\langle H \rangle)$, for 1000 realizations of the randomized careers. Fig. 3d-i shows three primary findings. First, before a hot streak, $\langle H \rangle$ is systematically larger than expected (z-scores $> 2$), indicating that individuals tend to diversify the topics they work on before a hot streak begins, consistent with an exploration strategy in the period leading up to hot streak. Second, following the onset of the hot streak, $\langle H \rangle$ measured in real careers becomes significantly smaller than expected (z-score $< -2$), suggesting that individuals become substantially more focused on what they work on, reflecting an exploitation strategy during hot streak. Third, despite the differences in the three types of careers we study and the methodologies to examine their career outputs, the observed associations between exploration, exploitation and hot streaks appear universal across all three domains we studied.

To systematically examine the temporal changes in entropy, we align careers based on when their hot streak begins and measure the dynamics of $H$ around hot streak (Fig. 3j-l). We find that compared with randomized careers, $H$ measured in real careers is systematically elevated before a hot streak begins, but drops precipitously below expectation during the hot streak. We further compare directly the entropy distribution $P(H)$ before and after the hot streak begins, finding that, across all three domains, $H$ during hot streak is systematically smaller than before (Fig. 3m-o, KS-test, p-value $< 0.001$); this pattern is absent when we repeat the same measurement for randomized careers (Fig. 3p-r).

The exploitation behavior during hot streaks appears consistent with several famous examples, including painter Jackson Pollock's "drip period" (1946-1950) (Fig. 2d), director Peter Jackson's *The Lord of the Rings* trilogy (Fig. 2h), and the career of scientist John Fenn, whose hot streak arrived late in his career but the work he produced during that period on electrospray ionization eventually won him the chemistry Nobel in 2002 (Fig. 2l). These examples raise an intriguing question: can the exploitation behavior by itself predict career hot streaks? To test this, we identify episodes of exploitation in each career by tracing the dynamics of $H$ across our three domains. We calculate the probability of initiating a hot streak with the onset of an exploitation episode, and compare it with the baseline probability measured in randomized careers (Fig. 3s-u). We find that when exploitation occurs by itself, not preceded by exploration, the chance that such episodes coincide with a hot streak is significantly lower than expected, not higher, across all three domains. These results indicate that exploitation by itself may not guarantee hot streaks, further suggesting the importance of prior exploration. Indeed, reexaminations of the careers of Jackson

Pollock, Peter Jackson, and John Fenn reveal a phase of unusual exploration of new and diverse art styles, types of films, and research topics, respectively, for the period leading up to their hot streaks (Fig. 2c, g, k). This observation raises the question of whether exploration that precedes a hot streak is instead the crucial ingredient, prompting us to calculate the probability of initiating a hot streak following an exploration episode alone. However, we find that when the episode of exploration is not followed by exploitation, the chance for such exploration to coincide with a hot streak again reduces significantly. By contrast, exploration followed by exploitation is consistently associated with a significant lift in the probability of initiating a hot streak: this configuration consistently outperforms the baseline across all three domains (20.5%, 13.8%, 19.2% over the baseline for artists, directors, and scientists, respectively), and represents the only positive lift among all combinations of the two creative strategies (Fig. 3s-u).

Taken together, these results demonstrate that neither exploration nor exploitation alone is associated with the hot streak dynamics; rather, it is the shift from exploration to exploitation that closely traces the onset of a hot streak. One plausible explanation is that exploration, as a risky, variance-enhancing strategy, increases one's chances to stumble upon new, potentially groundbreaking ideas; the subsequent exploitation behavior allows the individual to focus, develop knowledge and capabilities in that focal area, and build out their discoveries further. Importantly, our findings suggest that both ingredients of exploration and exploitation seem necessary. This supports the notion that not all explorations are fruitful, and that exploitation in the absence of promising new ideas may not be as productive. On the other hand, the sequence of exploration followed by exploitation may facilitate the emergence of high-impact work by incorporating new insights into a focused agenda. The positioning of exploration before exploitation may therefore serve to expand an individual's creative possibilities.

We test the robustness of our results across several dimensions. We split our samples of artists, directors and scientists based on the timing of their hot streaks (SI S3.1), the individual's level of impact (SI S3.2), and different fields of studies (SI S3.3), and repeat our analyses in each sub-sample, arriving at consistent conclusions. We further control for individual fixed effects in their exploration-exploitation dynamics (SI S3.4), and find that artists, directors and scientists predictably deviate from their typical creative behaviors around the beginning of a hot streak: individuals who tend to exploit become more exploratory before a hot streak begins, whereas individuals who tend to explore become particularly focused during their hot streak (SI S3.4). We further use regression analysis to fit the relationship between hot streaks and the exploration-exploitation transition by controlling for the impact of an individual's work, their career stage, and other individual characteristics, and find that our conclusions remain the same (SI S3.5).

For scientists who experience two hot streaks, we perform our measurements for the first and second hot streak separately (SI S3.6), and find that the exploration-exploitation dynamics hold true in both cases. And for those having hot streaks at the beginning of their careers, while by construction we cannot observe their prior behaviors, we find that they consistently engage in exploitation during their hot streaks (SI S3.7). We further verify that these results are robust to using different community detection algorithms such as Infomap[39] (SI S3.8) and different ways of aggregating data over time (SI S3.9). We also varied our entropy measure to quantify the exploration-exploitation dynamics by using a Simpson diversity measure $(1 - \Sigma_i p_i^2)$ (SI S3.10), also known as Gini impurity, and repeat all our results, finding again the same conclusions.

To understand the potential forces that might facilitate the shift from exploration to exploitation, we further examine the organization of innovative activity by focusing on scientific careers, asking whether there are detectable changes in collaboration patterns around the exploration-exploitation transition. We find that scientists are more likely to explore with small teams before a hot streak, but exploit with large teams after a hot streak begins. Indeed, we quantify the change in team size through two measures. We trace the dynamics of team size around the beginning of a hot streak (Fig. 4a). We also calculate the team size distribution observed in real careers normalized by the randomized careers (R(team size), Fig. 4b). Both results show that team size drops significantly before hot streak yet becomes substantially larger than expected during hot streak (Fig. 4a-b). These results are in line with the findings that small and large teams are differentially positioned for innovation[40]: large teams tend to excel at furthering existing ideas and design, whereas small teams tend to disrupt current ways of thinking with new ideas and opportunities. We further test the robustness of these results across different disciplines and control for the publication year, research field and career stage using regression analysis (SI S4), and arrive at the same conclusions.

Our final analysis probes potential connections between phases of exploration and exploitation surrounding a scientist's hot streak. We examine properties of the topics that are explored during the period leading up to hot streak, ranging from recency to citation impact to popularity, asking which ones tend to be chosen for subsequent exploitation. We find that the topic that was eventually exploited is less likely to be the one explored the most recently, or the highest cited, or the most popular among the topics explored before (see SI S5). These findings imply that, more than simply chasing after a discovery through exploration, individuals appear to seek out new opportunities by deliberating over different possibilities, and then harvesting promising directions through exploitation. To test if these potential connections can help us better understand which direction to exploit following exploration, we set up a simple prediction task to predict which topic to exploit using the features discussed above that characterize the exploration phase, including team size and topic properties (SI S5); this exercise yielded substantial predictive power (accuracy of 0.89 and AUC of 0.83). Overall, these results suggest intriguing connections between phases of exploration and exploitation surrounding a hot streak, which may have implications for science funding, especially given hot streaks and research grants tend to last for a similar duration.

Taken together, this paper reports among the first identifiable regularity underlying the onset of career hot streaks, which appears to apply universally across a wide range of creative domains. Overall, our results highlight the important role of both exploration and exploitation in individual careers. Curiously, across all three domains we studied, a major turning point for individual careers appears most closely linked with neither exploration nor exploitation behavior in isolation, but rather with the particular sequence of exploration followed by exploitation. Indeed, extant literature has documented the fundamental role of exploration and exploitation in creativity (SI S2.2, Table S1). Yet as creative behaviors, they have traditionally been considered either in isolation or in combination but rarely in succession[14,21]. Our results suggest a sequential view of creative strategies that balances experimentation and implementation may be particularly powerful for producing long-lasting contributions. These findings may hold broad relevance for identifying, training, and nurturing creative talents, especially given the various forces that sometimes appear in tension with the exploration-exploitation dynamics, ranging from the intensifying pressure to publish[41,42] to the specialization of individual expertise[10] and how such specialization is favored in personnel evaluations[43,44].

Note that, given the heterogeneity in career trajectories experienced across different individuals and creative domains, it is plausible that additional factors may also be at work. In this study, we tested several alternative explanations for the onset of hot streaks (see SI S6). Although each of these explanations we tested appears plausible by itself, we find that none of them show consistent associations, indicating that none of these alternative hypotheses alone can account for the hot streak dynamics we examined. It is also likely that on an individual basis, the exploration-exploitation transition is further influenced by other external factors[17,18]. Furthermore, individuals may receive short-term feedback (e.g. art critiques or peer reviews) that may offer additional signals shaping their career focus. Nevertheless, our results suggest that, despite the obvious heterogeneity in the settings we examined and the myriad factors that may affect career progression and success, the exploration-exploitation dynamics appears consistently associated with the onset of hot streaks across rather diverse domains.

Notably, the sequence of exploration followed by exploitation closely resembles strategies observed in a wide range of natural and socio-technical settings, from animal foraging[45] to human cognitive search[46], from multi-armed bandits and reinforcement learning[47] to role oscillation between brokerage and closure in social network[48]. It thus suggests that the sequential strategies of exploration followed by exploitation uncovered in this study may have broad relevance that goes beyond individuals' careers. Lastly, the representation techniques used in this paper could open up promising new avenues for research on creativity[36,49,50], offering a quantitative framework to probe the characteristics of the creative products themselves. Future advances in deep learning may enable researchers to incorporate more creative dimensions, and hence more fruitfully contribute to a computationally-enhanced understanding of creativity.

**Acknowledgement** The authors thank A.-L. Barabási, W. Ocasio, B. Uzzi, J. Evans, K. Rao, C. Candia, S. Medya, G. Tripodi, and all members of Center for Science of Science and Innovation (CSSI) for invaluable comments. This work is supported by Air Force Office of Scientific Research under award number FA9550-15-1-0162, FA9550-17-1-0089, and FA9550-19-1-0354.

**Author contributions** D.W. conceived the project and designed the experiments; L.L. and N.D. collected data and performed empirical analyses with help from J.C., C.L.G., and D.W.; all authors discussed and interpreted results; D.W., L.L., and N.D. wrote the manuscript; all authors edited the manuscript.

**Competing interests**: The authors declare no competing interests.

**Correspondence:** Correspondence and requests for materials should be addressed to D.W. (email: dashun.wang@northwestern.edu).

**Fig. 1. Quantifying individual creative trajectories using high-dimensional representation techniques**. **a**, The architecture of the deep neural network to build high-dimensional representation of artworks. We connect a pretrained VGGNet with three fully connected layers, and fine-tune the model with art styles labels. The blue box in the VGGNet indicates the convolutional layer and yellow box the max pooling layer. The green bar shows the top three styles predicted by the model for the input image (Self Portrait by Vincent van Gogh). We construct the high-dimensional representation of artwork by combining the output from the first and third convolutional layer (blue arrows) and the second fully connected layer (red arrow). **b,** An illustration of the 64 filters in the first convolutional layer. We highlight the first filter, the original image, and the output after the image passing the filter. The red box represents the size of the filter (3x3 pixel box). **c,** The activation of four layers in VGGNet and the saliency map of the post-impressionism class given the input image. The saliency map visualizes the important pixels that the model used to predict the post-impressionism art style. These maps show that the low-level layers capture the curvatures in the image (such as brushstrokes) whereas the high-level layers capture the shape of objects. **d,** Word embedding for film plots. Target words, such as 'gang' and 'men' in the plot of Dark Knight, are encoded as a binary vector and passed to the neural network. We use the hidden layer to represent the embedding of words and plots. **e,** Node embedding for the co-casting network. The toy example shows the co-casting network for major actors in five films. We apply DeepWalk to the co-casting network of 79K films, to capture the co-occurrence of nodes from the trajectories of random walkers. We use the hidden layer of the model to represent the cast information. We concatenate the word embedding from plots and the node embedding from casts to construct a 200-dimensional vector for each film. **f,** An illustration of the co-citing network among papers published by a scientist. Two papers are connected if they have at least one common reference, with link weight measuring the total number of references they share. Following prior work[16], we apply community detection algorithm to the co-citing network, and identify the topic of each paper as the community it belongs to.

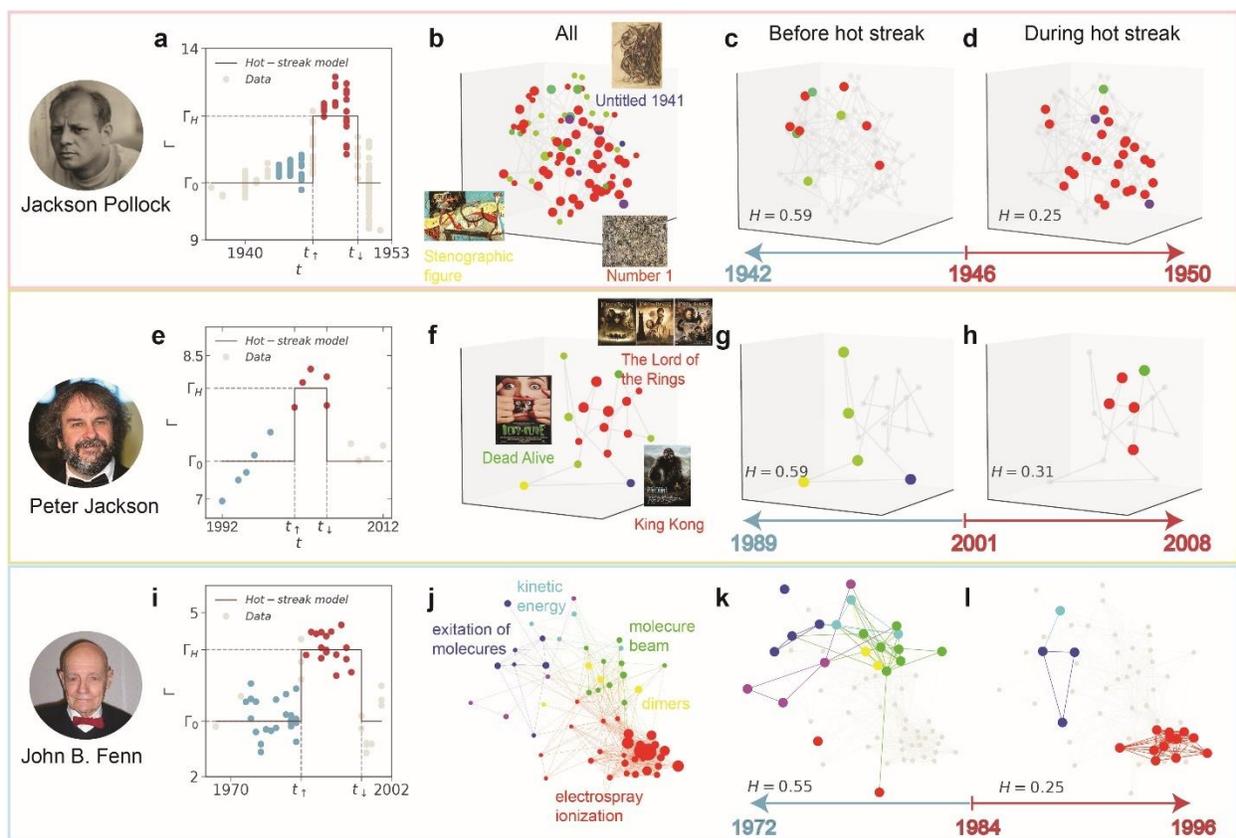

**Fig. 2 Creative trajectories and hot-streak dynamics: three exemplary careers. a, e, i**, We fit the hot streak model to careers of **a** Jackson Pollock, **e** Peter Jackson, and **i** John B. Fenn. The hot streak model[1] assumes that the impact of works produced in a career ($\log price$ for artworks, IMDB rating for films and $\log C_{10}$ for papers) is drawn from two normal distributions ($\Gamma_0$ and $\Gamma_H$), where $\Gamma_0$ captures the typical performance and $\Gamma_H$ captures the elevated performance during hot streak. We estimate the timing of hot streak by fitting the hot streak model (red line) to the impact sequence observed in real careers. $t_\uparrow$ and $t_\downarrow$ marks the beginning and the end of hot streak. To avoid mixing across the two periods, we measure the entropy of styles for artworks produced before (blue dots) and during hot streak (red dots) by excluding those produced during the year of the transition (excluding gray dots at the $t_\uparrow$ boundary). **b, c, d,** We project the 200-dimensional representation of artworks produced by Jackson Pollock to a 3D t-SNE embedding space. We apply k-means clustering to all images in our dataset (SI S1.2) and assign the cluster that each image belongs to as its style. Different styles are shown in different colors, and nodes with larger size denote those produced during hot streak. Jackson Pollock's hot streak is well aligned with the famous "drip period" (1946-1950). Indeed, the entropy of works produced during his hot streak is substantially lower than typical ($H = 0.25$ vs $H = 0.43$), suggesting an intensive focus on one particular style **d**. This exploitation behavior sharply contrasts the work he produced in the period leading up to hot

streak, which was characterized by an unusual exploration of new and diverse styles ($H = 0.59$) **c. f, g, h,** We project the 200-dimensional representation of films produced by Peter Jackson to a 3D t-SNE embedding space. Similarly, we apply k-means clustering to all films in our dataset and use clusters to approximate different styles (different color). Node size correlates with the IMDB rating of the film. To calculate the entropy of styles, we use all works Jackson produced during hot streak (red dots), and the same number of works before the hot streak (blue dots). Peter Jackson's hot streak covers *The Lord of the Rings* trilogy ($H = 0.31$) **h**. Before his hot streak, however, Jackson worked on diverse types of films including biography and horror comedy ($H = 0.59$) **f. j, k, l,** For the career of John Fenn, we study the co-citing network of his papers, with the communities within this network capturing different research topics he worked (color coded). Node size correlates with a paper's impact. Before his hot streak, Fenn worked on numerous different topics from excitation on hot surfaces to dimers ($H = 0.55$) **k**. But during his hot streak, Fenn intensively focused on electrospray ionization ($H = 0.25$) **l**, which eventually won him the chemistry Nobel in 2002.

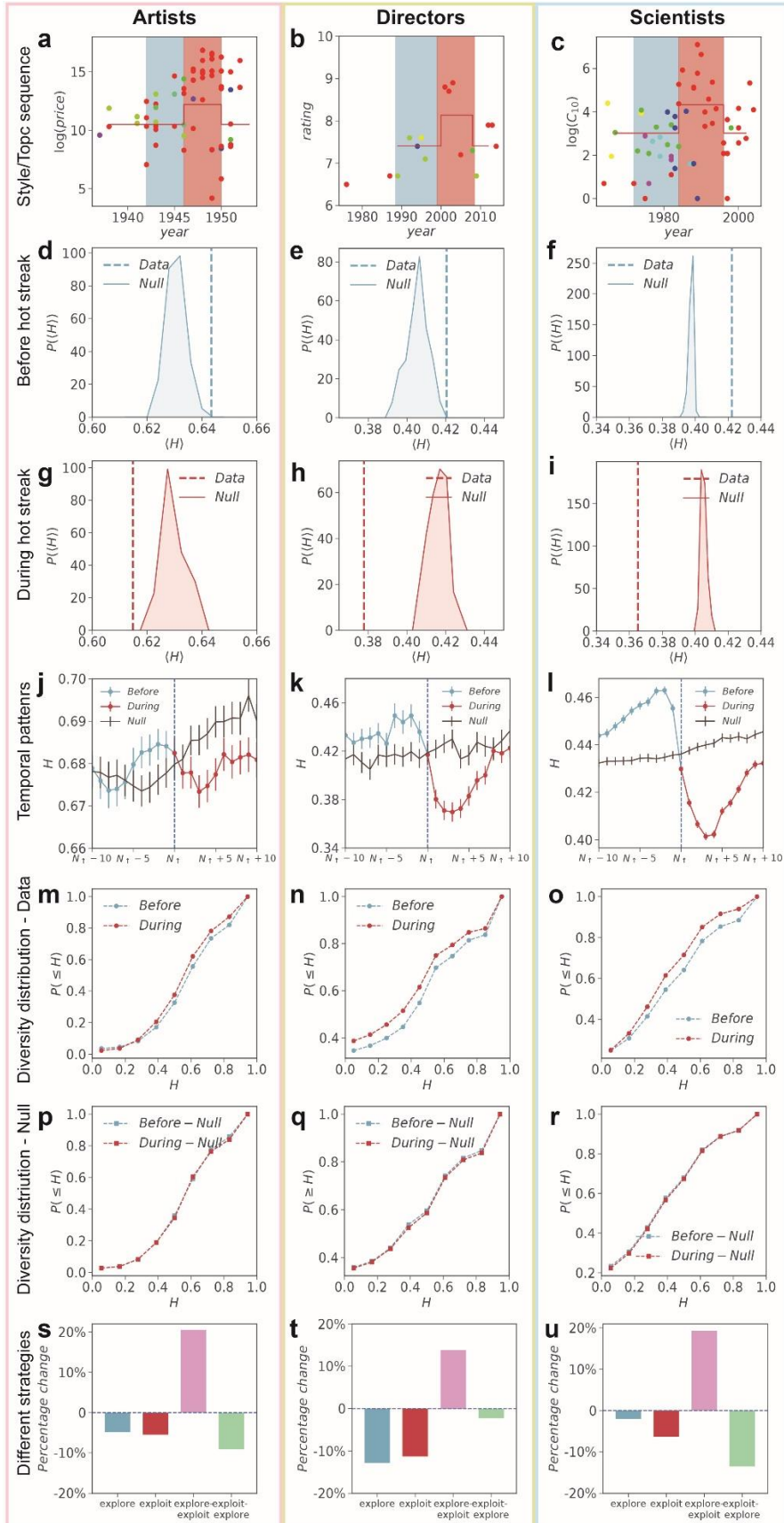

**Fig. 3 Exploration, exploitation and career hot streaks. a, b, c**, Career histories of **a** Jackson Pollock, **b** Peter Jackson, and **c** John Fenn illustrate the topics they worked on before and during their hot streak and the impacts of the work. Color of the dots is consistent with the dots shown in Fig. 2**bfj**. **d, e, f,** The distribution of entropy $P(\langle H \rangle)$ before a hot streak for 1000 realizations of the randomized careers for all individuals analyzed in our datasets. The vertical line indicates $\langle H \rangle$ measured in real careers, showing that it is significantly larger than expected (z scores are 4.24 for artists, 2.94 for directors, and 13.90 for scientists.) **g, h, i,** Same as D-F but for the entropy of work produced during hot streak. $\langle H \rangle$ in real careers (vertical line) is significantly smaller than expected (z scores are $-2.42$ for artists, $-8.54$ for directors, and $-22.71$ for scientists.) **j, k, l,** The dynamics of topic entropy $H$ surrounding the onset of hot streak for real and randomized careers, measured through a sliding window of six artworks, five films or five scientific papers. Error bars represent the standard error of the mean. **m, n, o,** Cumulative entropy distribution $P_{\leq}(H)$ before and during hot streak in real careers across the three domains. *P*-values of the KS-test are $3.7 \times 10^{-6}$ for artists, $1.5 \times 10^{-5}$ for directors, and $1.1 \times 10^{-64}$ for scientists. **p, q, r,** Cumulative entropy distribution $P_{\leq}(H)$ before and during hot streak for the null model. *P*-values are 0.23 for artists, 0.77 for directors, and 0.06 for scientists. **s, t, u,** The probability to observe the onset of hot streak at the end of an exploration episode alone (not followed by exploitation), or at the beginning of an exploitation episode alone (not proceeded by exploration), or at the transition from exploration to exploitation, or from exploitation to exploration. We then compare with the baseline probability of having a hot streak. Here we calculate entropy with a sliding window of two years for artists and scientist, and five works for directors, and define exploration and exploitation episodes as entropy above or below one's average.

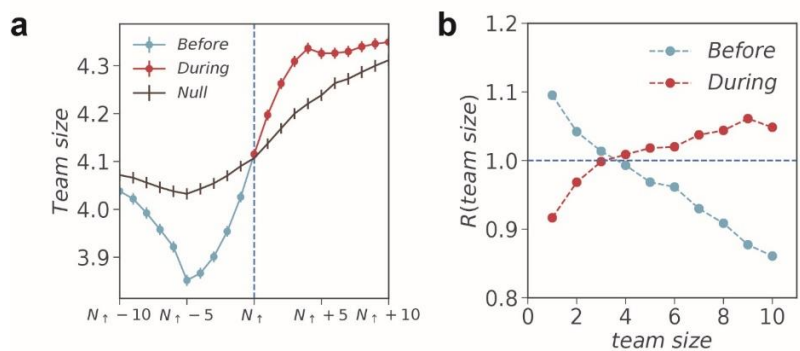

**Fig. 4 Authorship structure and hot streaks in science**. **a**, The average team size around the beginning of a hot streak for real and randomized careers in science. Team size shows a significant drop before a hot streak begins but a notable increase during hot streak. **b**, We calculate team size for papers published before and during hot streak, and compare the distribution to that of randomized careers, denoted as *R*(team size). *R* decreases with team size from above to below 1 for papers published before a hot streak but increases with team size for papers published during a hot streak. Both measures in A and B suggest that scientists tend to engage with smaller teams before hot streak, and with larger teams during hot streak.

# Supplementary information for understanding the onset of hot streaks across artistic, cultural, and scientific careers

Lu Liu,[1, 2, 3, 4] Nima Dehmamy,[1, 2, 3] Jillian Chown,[1, 3]
C. Lee Giles,[4, 5] and Dashun Wang[1, 2, 3, 6, *]

[1]*Center for Science of Science and Innovation,*
*Northwestern University, Evanston, IL, USA.*
[2]*Northwestern Institute on Complex Systems,*
*Northwestern University, Evanston, IL, USA.*
[3]*Kellogg School of Management, Northwestern University, Evanston, IL, USA*
[4]*College of Information Sciences and Technology,*
*Pennsylvania State University, University Park, PA, USA.*
[5]*Department of Computer Science and Engineering,*
*Pennsylvania State University, University Park, PA, USA.*
[6]*McCormick School of Engineering,*
*Northwestern University, Evanston, IL, USA.*

* Corresponding author: dashun.wang@northwestern.edu

**CONTENTS**

## S1. DATA AND METHOD

In this paper, we constructed large-scale datasets for creative products from a wide range of disparate sources, including images of artworks for artists, film plots and casts for directors, and publication and citation information for scientists, along with their impact measures of auction prices, IMDB ratings and paper citations, respectively [1]. We developed computational tools from deep learning and network science and learned high-dimensional representations for these creative products, allowing us to trace an individual's career trajectory on the underlying creative space around the beginning of a hot streak. In this section, we describe in detail how we construct the data and how we learn the representation for the three domains.

### S1.1. Artists

*S1.1.1. Data*

We collect for artists large-scale image datasets tracing art styles, and auction datasets tracing the career impact, to quantitatively understand artistic success through the work one produced. We first gather over 800K images of visual arts from two databases, Art500K [2, 3] and Artnet (www.artnet.com). Art500K contains large-scale images of artworks collected from several online sources and museum collections such as Rijks Museum and Wikiart, covering artworks for over 600 years from medieval to contemporary art. The data contain information regarding disambiguated artist names, title, year of production for each artwork. In addition, Art500K offers art style labels for over 160K images, allowing us to train the art style embedding in the classification task. Artnet is an art market website with over 340K disambiguated artist profiles. The free version of the website records for each artist a list of gallery collections and auction records in the market, along with the image, title, and production year. If an artwork was produced over a span of several years, we use the last year as its production year, corresponding to the year in which the work was finalized. Together, we curated for artists in Art500K and Artnet their image profiles with the list of images and year of production. To quantify the impact dynamics and career hot streaks, we further collect impact profiles of 3,480 artists who have at least 15 artworks and 10 years of career length by combining two art market databases, Artprice (www.artprice.com) and Findartinfo (www.findartinfo.com). Building on prior work [1, 4], we curated for each artist the list of auction price and year of production for the work one produced, allowing us to measure the timing of hot streak for each individual with the auction price.

We compare each of the 3,480 artists who have impact information to the image profiles curated from Art500K and Artnet. Two artists are considered to be the same if they satisfy the following criteria: 1) Identical names. The last name and first initial are the same. If full names are available for both artists, they must be identical. 2) Active during the same period. The career span of the image profile, measured by the production year of the first and last artwork, is within 10 years of the career span of the impact profile. By applying this entity linkage procedure, we end up with 2,016 artists with impact sequence and at least 10 images for our analysis. Note that images from the Art500K mainly include artworks from museums that may not appear in the auction market. Previous studies show that the years during which artists produced artworks with high market value are aligned with the production years of high-value artworks considered by museum curators [4, 5], suggesting that the timing when an artwork was produced is useful to identify its relative impact within a career.

Although our datasets capture among the largest collections of career profiles, there are limitations of the data that readers should keep in mind. For example, the coverage of one's entire career history may vary across individuals, and famous artists may have a better coverage overall than their less famous counterparts. Artworks by influential artists are more likely to appear in the art market or in the museum collection, whereas for less famous artists, the museums and art markets may only include their best work. Similarly, the coverage of artworks within each career might not be uniform, possibly biased toward the famous work one produced. This could lead to a potential downward bias for the exploration phase before the hot streak but a potential upward bias for the exploitation results during the hot streak.

*S1.1.2. Method*

We train an art style classifier with deep neural networks using images with ground-truth labels in Art500K [6]. If an image have multiple labels, we randomly pick one as its unique label. Art500K provides 230 unique art styles, with imbalanced sample size among them (Fig. S1). To improve the performance of the classifier, we ignore styles with too few images, and reduced the number of classes to art styles with over 1000 images (top 29). For classes with more than 5000 samples, we randomly selected 5000 images for each class. For classes with samples fewer than 5000, we increase their sample size to 5000 by applying image augmentation methods with rotation (±30), scale (1-1.2) and random crop (0.8-1). Together, we gathered a balanced dataset of 145K images for the classification task. We normalize each image by their mean and standard deviation. We randomly sample 80% of images as the training set and the rest 20% as the test set.

We build on a popular convolutional neural network architecture in art style analysis, VGG16 [7]. Prior studies show that a transfer learning approach in which a model is fine-tuned outperforms the model trained from scratch, as its filters are more interpretive and better capture lines and shapes learned from the object recognition tasks [6], prompting us to adopt the transfer learning method in this paper. Specifically, we remove the dense layers in the pretrained VGG16, added two hidden dense layers with ReLU activation (dimension = 4096 and 512, respectively) and a final classification layer of 29 nodes with softmax activation (Fig. 1A). The first dense layer learns the non-linear combinations of outputs from convolutional layers. The second dense layer further reduces the dimension for the output of the previous layer and learns the high-level content for each image to better identify different art styles. In the experiment, we fix the parameters of the convolutional layers and only fine-tune the weights in the three newly added layers. We train the model with Adam optimizer and batch of size 128, and selected the model with the best performance. The highest accuracy is 0.54 on the test set, which is significantly higher than the random accuracy for our balanced set (0.03). A recent study on art style classification [6] utilized 77K images with art style labels from Wikiart, and trained their model with unbalanced samples from 20 style classes. Their random accuracy is 0.17 and their best accuracy is 0.637, suggesting that the performance of our model is comparable to the state-of-art in terms of the accuracy improvement compared to random guess.

As art style is encoded in both low-level features like brush strokes and high-level features like content and themes [8], we combine both the low- and high-level embedding to represent the art style of each image. We first use the 512-dimensional dense layer to capture the high-level feature of each image. We extract the output from the 512-dimensional dense layer for all images in our dataset, and then apply principle component analysis (PCA) to reduce the 512-dimensional vector to a 100-dimensional vector. We further generate low-level features from the first and third convolutional layers in VGG16 ($d$ = 224 × 224 × 64, $d$ = 112 × 112 × 128, respectively). Different from the traditional Gram matrix kernel [2, 8] that flattens the output of convolutional layers to generate a huge vector, here we reduce the dimension of the two tensors by calculating the mean value for the second and third dimension, and generate a 224-dimensional vector and a 112-dimensional vector respectively. As the absolute value of the two vectors may not be comparable to compute the distance in the embedding space, we further normalize the two vectors by their sample mean and standard deviation, and then concatenate the 224-dimensional vector from the first convolutional layer and the 112-dimensional vector from the third convolutional layer. To make sure that the low- and high-level features are of equal importance, we further reduce the low-level feature to 100-dimensional vector by applying PCA on the 336-dimensional vector. In the end, we normalize the low and high level embedding by their sample mean and standard deviation, and combine the two 100-dimensional vectors together to get a 200-dimensional representation for each image. We visualize the kernel density projected onto two dimensions with PCA for images from three art styles (Fig. S7). We then apply the k-means clustering to the 200-dimensional embeddings, and assigned style label for each image as the cluster it belongs to. In the main text we report results based on k-means clustering with $n$ = 30 centroids. We show in Fig. S29 that our results are robust to different number of centroids.

To better understand the meaning of each layer, and to test if the selected layers can indeed capture low-level brushes strokes and high-level content of an image, we offer several case studies on famous paintings to obtain more insights from the fine-tuned model. We analyze two positive samples, Number 1 by Jackson Pollock (Fig. S3) and Mona Lisa by Leonardo da Vinci (Fig. S4), and two negative samples, Kiss by Gustav Klimt (Fig. S5) and Café terrace by Vincent van Gogh (Fig. S6). Our model can successfully predict the abstraction expressionism for Number 1 and high renaissance for Mona Lisa, while it misclassifies Kiss, one of the most famous paintings in Art Nouveau, and Café terrace, a famous post-impressionism painting, to Surrealism. We visualize a random activation from the first and third convolutional layer for each image. The first convolutional layer learns to detect lines, dots and colors from raw pixels with 3 × 3 filters, and then passes these local patterns to deeper layers, which captures the brush strokes within a broader region of each image. For example, the dripping and splattering in Jackson Pollock's Number 1 are well captured by these low-level layers, and the uniform distribution of each pattern also reflect the balanced nature of his work. Compared to the local patterns coded in low-level layers, the activation of the convolutional layer in the last block and the last max-pooling layer show higher-level features such as the landscape, part of a body, and the roof of a house, which suggests that high-level features capture more about the content or object in an image. Finally, we visualize the saliency map for the predicted label, which highlights the important pixels that the model utilized to make the decision. The two negative samples share common features in the saliency map, where the yellow regions in the background and the clothes for Kiss, and the yellow cafe house for Café terrace are highlighted, which illustrates why the two images may be classified to the same style.

## S1.2. Directors

### S1.2.1. Data

Our second setting traces the profiles of film directors from The Internet Movie Datavase (IMDB) (www.imdb.com/interfaces). The IMDB database contains one of the largest collections of film records for over 100 years worldwide, along with detailed information for each film such as title, release time, user rating and more. In addition, the webpage for each film includes sections for full cast and crew list, plot information, and genre tags, allowing us to learn the embedding of films. The IMDB database also provides a disambiguated profile for each director with a unique ID assigned to the list of work they produced. We focus on directors who have at least 10 years in career length and 15 films in IMDB. We use the IMDB rating to approximate the impact of each film and measure the timing of hot streak within a career with the rating sequence [1, 9]. For films produced by the directors in our dataset, we further collect the plot and cast information from their IMDB webpages. Specifically, we gather plot information from the storyline, plot summary and plot synopsis section. If a film has multiple paragraphs from these sections, we merge them together to construct a single plot record for each film. We also gather for each film the full list of actors with their unique IDs. Given IDs in the IMDB database, we can identify if a film belongs to the hot streak period or not. In total, we collect 79K films by 4,377 directors for our analysis.

*S1.2.2. Method*

Prior studies show that the storyline and casting information play an important role in the success of a film [10–14], prompting us to construct for each film the word embedding from plot summaries and node embedding from co-casting network. We first preprocess the plot text to learn the word embedding. We tokenize each plot into words and transferred all words to the lowercase. We remove punctuation and English stop words, stem each word, and only keep stemmed tokens with at least 3 characters. We train the word embedding with skip-gram model in our plot corpus from scratch, which captures the meaning of a word from its surrounding text in the plot corpus. We learn the 100-dimensional representation for each word with window size of 5 for 50 iterations, and calculate the plot embedding vector as the average value for all word vectors in each plot.

We further incorporate the casting information by constructing a co-casting network for the work produced by directors in our dataset. As IMDB ordered the cast by their importance, here we focus on top 10 featured actors. Two films are connected if they have common actors, with the link weight as the total number of actors they shared. We apply DeepWalk [15] to the co-casting network, a well-known representation algorithm for social networks which captures the local structures of nodes from random walks. We set the walk length as 10, the number of walks as 80, window size as 5, and learn the 100-dimensional node embedding. We further normalize the 100-dimensional plot and cast vector by sample mean and standard deviation, and concatenate the two vectors to get a 200-dimensional vector for each film. We visualize the kernel density for films from three different genre projected onto the two dimensional embedding space with PCA (Fig. S8). Similarly, we apply the k-means clustering to the 200-dimensional embeddings with their euclidean distance, and assign the style label for each film as the cluster it belongs to. In the main text we report results based on k-means clustering with $n = 30$ centroids. We show in Fig. S29 that our results are robust to different number of centroids.

To test if the learned embedding captures the film styles, we use the 200-dimensional vector as the input to predict the genre of each film. We formulate this task as a multi-label classification problem. Specifically, we build a fully-connected neural network with two hidden layers (100d and 50d). We randomly split the data into training (80%) and test set (20%), and use the Adam optimizer to train the model with batch size of 128 for 20 epochs. Although we do not utilize any genre information to learn the film embeddings, they can successfully predict film genres with an accuracy of 0.948 (Fig. S9A). We repeat the classification task for word embedding and cast embedding separately (Fig. S9B-C), and find that they can also achieve a high accuracy (0.94 for both cases). And combing the two leads to even better performance.

### S1.3. Scientists

*S1.3.1. Data*

In our third setting, we analyze publication records of scientists by combining Google Scholar (GS) profiles with Web of Science (WoS) [1]. GS allows individual scientists from diverse disciplines to create, maintain, and update their publication profiles. Assisted by its disambiguation algorithms, GS profiles provide a state-of-art disambiguation method to assemble publication list for individual scientists, offering unique opportunities to study scientific careers [16, 17]. We collect scientists' profiles and matched each publication record to the WoS database, which provides publication metadata and citation records for around 46 million journal papers since 1900. We curate for each scientist a list of papers with unique WoS identification numbers, publication date, and citation in 10 years $C_{10}$ to approximate the impact of each paper, yielding 20,040 scientists with at least 15 papers and 20 years of career length for our analysis. We measure the timing of hot streak for these scientists using the logarithmic of $C_{10}$ for the sequence of work they produced [1].

*S1.3.2. Method*

We identify the topics of over 1 million papers published by these scientists using a novel network method introduced by Zeng and colleagues [18]. Here, we extract the reference list of each paper from the WoS with unique WoS ID, and focus on papers with at least one reference as recorded in the WoS. We construct a weighted co-citing network among papers produced by each scientist. Two papers are linked if they have common references, with the link weight as the total number of references they shared. We apply community detection algorithm to the ego co-citing network for each scientist and assign the topic of a paper as the community it belongs to. In the main text, we report results using the community detection methods. To examine the robustness of our results, next we follow the same methodology for artists and directors to learn an embedding for papers with neural networks.

We construct a co-citing network among 1 million papers published by individuals in our dataset. Two papers are linked if they have common references, with the weight indicating the number of references they share. To speed up the network embedding algorithm, we first reduced the size of its adjacency matrix $A$ with truncated SVD to 500 dimensions, where $A = U^T \Sigma V$. Fig. S10 shows the cumulative variance explained. There is no obvious saturation after a certain $n$, suggesting all components contain important information for the matrix. We then use the subject label in WoS to train a neural network and learn the representation of each paper. We focus on the top 34 subjects with more than 10,000 papers in the dataset, and ignore subjects with too few papers in this case. We randomly sample 10,000 papers for each subject if it contains more papers to make a balanced training set. We randomly split the data into training (80%) and test set (20%). We train a neural network with one hidden layer (64d). The accuracy is 0.515 for the test set (Fig. S12). We use the 64-dimensional vector from the hidden layer to represent the topic information of a paper. We visualize the kernel density for papers from three different disciplines in the embedding space (Fig. S13). We apply k-means clustering to the 64-dimensional embedding space with $n = 60$ centroids, and assign the topic of each paper as the community it belongs to. We repeat our analysis on *(H)* for topics measured from the embedding space, finding our conclusions remain the same (Fig. S14).

## S2. RELATED WORK

### S2.1. Art style analysis

In this section, we briefly summarize related work on art image analysis from three directions: 1) statistical patterns of art styles, 2) deep neural networks (DNN) and style classification, and 3) neural style transfer.

**Statistical patterns of art styles** This line of research focuses on providing quantifiable information on art styles by looking at different visual properties of the images [19]. Researchers proposed various metrics from local patterns of images to identify art styles or artists, ranging from the fractal dimensions [20–23] to edge and shape statistics [24] to color usage [25–27]. Early studies mainly involve with small samples, consisting of several thousand images or less, and focus on case studies of certain art styles or artists [20, 22, 24]. More recent work started to cover images from different art movements thanks to the development of digital libraries on artworks such as Web Gallery of Art and Wikiart. For example, Sigaki *et al.* analyzed art history from images in Wikiart [28], and measured the entropy and complexity of local image patterns, and found identifiable trend of art style evolution in the complexity–entropy plane. Although this line of research provides insights to quantitatively understand art styles, researchers usually need to design meaningful local features that can better identify different art styles, the process of which is largely accelerated by the DNN studies that can automatically learn image representations.

**DNN and style classification** One productive line of research uses DNN to learn a better representation of images with large-scale labeled images of visual arts. Wikiart dataset is a popular dataset for such tasks [6, 29–33]. In addition, more recent studies curated large-scale, structured datasets on images of visual art by combining different online sources, and increased the number of art images to over 500k [2, 3, 34]. These studies built upon popular DNN architectures such as VGGNet [7], ResNet [35] and AlexNet [36] in object recognition, and fine-tuned models pre-trained on ImageNet or trained these models from scratch [2, 6, 29, 31–33, 37]. Although DNN improves the accuracy of art style classification with high-level representations, researchers find that the joint embedding from both high-level vector and the output of convolutional filters better captures the art styles [2], and adding outputs from convolutional layers to the high-level embedding also outperforms the high-level embedding in style and artist classification [30, 31], prompting us to use the joint embedding for style representation in this paper.

**Neural style transfer** Another line of research examines the art style analysis using neural style transfer [38]. The seminal work by Gatys *et al.* [8] first proposed to use VGGNet to convert an natural image into the art style of a target painting by optimizing the content information of the original image coded in high-level filters and the art style of the target painting coded in low-level filters. Since then researchers have studied various methods to speed up the transfer process [39, 40] and apply this method to different scenarios (e.g. general-purpose image-to-image transfer) [41] and different types of inputs (e.g. videos) [42]. More recent studies also use generative adversarial networks (GANs) for style transfer [43], and extend the input from a single image to an embedding space of styles. GAN has also been used to generate creative artworks that deviate from the trained distributions [44].

### S2.2. Exploration and exploitation

The trade-off between exploration and exploitation — and its relationship to creativity and learning — has been discussed extensively across a broad set of disciplines, ranging from computer science [45–47], to psychology [48, 49], to neuroscience [50, 51], to computational social science [52–54], to strategic management and organization theory [55–59]. On the one hand, producing creative and high-impact works requires one to explore new and diverse ideas, especially given the combinatorial nature of innovation and technology. Yet, exploration often comes at the cost of productivity. Exploitation, on the other hand, can support a focused agenda which is essential for developing existing knowledge. The trade-off between exploration and exploitation represents an enduring dilemma for individual and organization learning, motivating a large body of literature to examine exploration and exploitation from both theoretical and empirical perspectives. In Table. S1, we offer a brief overview of three relevant lines of research inquiring individual behavior, organization learning, and idea formation. We next discuss these directions in more detail.

**Individual behavior** At the individual level, the 'essential tension' hypothesis by Thomas Kuhn [60] illustrates the choice between exploiting existing ideas and exploring new yet risky opportunities. The sociology of science offers several fundamental theoretical discussions [61, 62]. More recently, empirical anlaysis has been conducted to quantitatively understand the 'essential tension' hypothesis. For example, Foster *et al.* [53] analyzed millions of abstracts from MEDLINE, and identified topics from the clusters on the chemical network to trace the research strategy of biomedical researchers [63]. In addition, the PACS code in American Physical Society (APS) dataset has also been widely used to quantify exploration and exploitation for scientific careers [18, 52, 64].

Researchers have also studied various environmental, social and individual factors that may influence one's choice between exploration and exploitation [48]. Environmental factors include resource status of a local position [49, 65], cost and reward of exploration and exploitation [65, 66], available information on different options [67], and more. Discussions centered around how long individuals should stay in the exploitation/exploration phase and when to change their behaviors under different environmental settings. For example, the probability of exploration increases when the resource is depleted, when the cost of exploration decreases, or when individuals are uncertain about the options. The social factors are widely discussed in social learning strategies and collective intelligence [68–72], ranging from task complexity [73], to past success and failure [71, 73] to network structures [74, 75]. Individuals can update their strategies like exploration, exploitation or copying others to increase their payoffs under different settings. Individual factors such as personalities [76], cognitive capacity [77], and aspiration level [78], also influence one's propensity to explore or exploit.

In the literature of strategic management and organization theory, scholars have examined exploration and exploitation behaviors of individuals and firms, particularly focusing on the effects this has on organizational outcomes. For example, Singh & Agrawal [79] found that when scientists begin working within a new organization, the organization increases their use of the new recruit's prior work and that the majority of the effect is due to the employee's own exploitation of their prior work. Groysberg & Lee [80] found that when star security analysts were hired to explore (i.e., to initiate new activities for the organi-zation), they experienced a drop in performance; whereas star security analysts hired to engage in exploitation (i.e., to reinforce the organization's existing activities) experienced a boost in performance. Other research has looked at the antecedents of individuals' exploration and exploitation behaviors. For example, Lee & Meyer-Doyle [81] examined how financial incentives shaped the behavior of sales people and found that individuals engaged in more exploration when performance-based incentives were weakened but this increase was driven by the organization's strongest performers. Recent study on network oscillation for bankers [82] suggests that switching between exploration and exploitation has positive effects on the employee's network advantage.

**Organization learning, design and adaptation** At the macro level, another important line of research examines exploration and exploitation in the context of organization learning, organization design, and organizational adaptation [58]. This line of work builds on the canonical work by March [57], and suggests that both exploration and exploitation are critical for an organization's performance, but they are inherently in tension and that this tension must be actively managed [83]. This tension reflects trade-offs between short vs. long-term performance and stability vs. adaptability [57, 84–87]. Debates in this literature center on several fundamental questions: Do exploration and exploitation exist as two ends of a continuum (and so cannot coexist at the same time) or are they orthogonal discrete choices? Can organizations find a balance between exploration and exploitation activities or should they specialize in one or the other? It also explores the antecedents to organizations' decisions to pursue exploration or exploitation [59, 88], examining environmental factors (e.g., exogenous shocks, competitive dynamics) as well as organizational factors (e.g., culture, resources, capabilities) that influence that choice. This literature also uses the notion of organizational ambidexterity to describe the ability to do both exploration and exploitation simultaneously [89]. Finally, this research examines the performance implications for organizations of adopting different approaches to balancing this enduring trade-off between exploration and exploitation [90]. This line of research is performed using multiple different methodologies including empirical studies using quantitative and qualitative data from organizations, theoretical models [91], and agent-based simulations [59, 92, 93].

**Idea formation** At a more micro level, the discussion of exploration and exploitation is particularly relevant to studies on idea formation and innovation process [94–96], which models the mechanism of innovation as random walks on the network of ideas/landscape of solutions. In this setting, exploration and exploitation is usually defined as creating new path or reproducing existing ideas. For example, Iacopini et al [94] models the cognitive growth of knowledge in science for over 20 years and validate process with concept networks curated from WoS abstracts. Studies have shown that both existing knowledge and novel combinations are essential for producing high-impact scientific papers [97]. The discussion goes beyond science to innovation and technology as well. For example, Youn *et al.* [98] analyzed technology codes used by USPTO to quantify innovation strategy, finding a constant rate of exploration and exploitation in patent records.

Overall, our results contribute to these three lines of literature in several ways. First, by documenting the relationship between exploration, exploitation and career hot streaks, our results demonstrate broader relevance of the concepts of exploration and exploitation, extending beyond existing individual or organizational settings to the understanding of hot streaks and individual creative careers. At root, our results suggest the important role of both exploration and exploitation in individual careers. Curiously, across a wide range of creative domains, a major turning point for individual careers appears most closely linked with neither exploration nor exploitation behavior in isolation, but rather with the particular sequence of exploration followed by exploitation, which highlights our second contribution. Indeed, extant literature has documented the fundamental role of exploration and exploitation in creativity. Yet as creative behaviors, they have traditionally been considered either in isolation [53, 60] or in combination [58, 99] but rarely in succession. Our results suggest a sequential view of creative strategies that balance experimentation and implementation may be particularly powerful for producing long-lasting contributions.

| Year | Paper | Category | Topic |
|------|-------|----------|-------|
| 2019 | Aleta *et al.* [52] | Individual | topic evolution for scientific careers |
| 2019 | Shibayama *et al.* [100] | Individual | junior scientists' performance and their training strategy |
| 2019 | Zeng *et al.* [18] | Individual | topic evolution for scientific careers |
| 2018 | Iacopini *et al.* [94] | Idea | emergence of knowledge and innovation |
| 2018 | De Langhe *et al.* [101] | Individual | exploration, exploitation and scientific revolution |
| 2018 | Luger *et al.* [86] | Organization | dynamic balancing of exploration and exploitation |
| 2017 | Jia *et al.* [64] | Individual | topic evolution for scientific careers |
| 2016 | Loreto *et al.* [95] | Idea | emergence of knowledge and innovation |
| 2016 | Piao & Zajac *et al.* [87] | Organization | repetitive versus incremental exploitation and impact on exploration |
| 2016 | Murdock *et al.* | Individual | the reading strategy of Charles Darwin |
| 2016 | Burt & Merluzzi [82] | Individual | oscillation of network structure between brokerage and closure |
| 2015 | Foster *et al.* [53] | Individual | essential tension in biomedical research |
| 2015 | Rzhetsky *et al.* [63] | Idea | the evolution of problem selection in in biomedical research |
| 2015 | Youn *et al.* [98] | Idea | combinatorial dynamic of exploration and exploitation in innovation |
| 2015 | Bateman & Hess [76] | Individual | scientists' personality and research strategy |
| 2015 | Krafft *et al.* [71] | Individual | collective intelligence and learning strategies |
| 2015 | Chin *et al.* [102] | Individual | age and search strategy |
| 2015 | De Langhe & Rubbens [103] | Idea | essential tension in science |
| 2014 | Holmqvist [84] | Organization | exploitation and exploration in inter-organizational learning |
| 2014 | Knudsen & Srikanth [93] | Organization | coordinated behaviors by multiple individuals |
| 2014 | Tria *et al.* [96] | Idea | dynamics of correlated novelties |
| 2014 | Spisak *et al.* [104] | Individual | age of leaders and search strategy |
| 2014 | Billinger *et al.* [73] | Individual | impact of task complexity on the trade-off |
| 2014 | Toyokawa *et al.* [70] | Individual | impact of social learning on the trade-off |
| 2013 | Uzzi *et al.* [97] | Idea | novelty and conventionality of knowledge and paper impact |
| 2013 | Berger-Tal *et al.* [54] | Individual | trade-off for project-levels and project-based strategies |
| 2012 | Hills *et al.* [77] | Individual | memory and individual searching strategies |
| 2012 | Posen & Levinthal [59] | Organization | exploitation and exploration in inter-organizational learning |
| 2011 | Molina *et al.* [105] | Organizations | goal of new product development and the strategy choice |
| 2011 | Singh & Agrawal [79] | Individual | inventors mobility and exploration/exploitation behaviors |
| 2010 | Fang *et al.* [106] | Organization | balance between exploration and exploitation |
| 2010 | Rakow & Newell [67] | Individual | available information and propensity to take risks |
| 2009 | Groysberg & Lee [80] | Individual | individual mobility and exploration/exploitation behaviors |
| 2009 | Weisberg & Muldoon [107] | Individual | search strategy on epistemic landscape |
| 2008 | Hau *et al.* [66] | Individual | the role of sample size on decision making |
| 2008 | Goldstone *et al.* [72] | Individual | collective behavior and different learning strategies |
| 2007 | Namara *et al.* [108] | Organization | trade-off in biotechnology firms |
| 2007 | Eliassen *et al.* [109] | Individual | lifetime expectancy and individual foraging strategies |
| 2007 | Sidhu *et al.* [110] | Organization | multidimensional search in supply, demand, and geographic space |
| 2007 | Lazer & Friedman [75] | Individual | network structure of exploration and exploitation |
| 2007 | Lee & Meyer-Doyle [81] | Individual | incentives and the behavior of sales people |
| 2007 | Parker *et al.* [78] | Individual | maximizers, satisficers, and preference to exploration |
| 2006 | Lavie & Rosenkopf [111] | Organization | balancing exploration and exploitation in alliance formation |
| 2006 | Siggelkow & Rivkin [92] | Organization | performance effects of balance between exploration and exploitation |
| 2005 | Jansen *et al.* [88] | Organization | ambidexterity of exploration and exploitation |
| 2005 | Auh & Menguc [90] | Organization | multiplicative interactions between exploration and exploitation |
| 2004 | Rothaermel & Deeds [112] | Organization | exploration and exploitation in alliance formation |
| 2004 | He & Wong [113] | Organization | ambidexterity of exploration and exploitation |
| 2002 | Burgelman [114] | Organization | long-term adaptive capability of a firm's strategy |
| 2001 | Rosenkopf & Nerkar [56] | Organization | patenting activity and the external environment |
| 2001 | Sørensen *et al.* [85] | Organization | organizational aging and innovation process |
| 1994 | Henderson & Cockburn [55] | Organization | R&D behaviors and research productivity in pharmaceutical firms |
| 1993 | Levinthal & March [115] | Organization | exploration and exploitation in organizational learning |
| 1991 | March [57] | Organization | exploration and exploitation in organizational learning |
| 1979 | Kuhn [60] | Individual | essential tension in science |
| 1976 | Charnov [65] | Individual | optimal foraging behaviors |
| 1975 | Bourdieu [61] | Individual | scientific organizing framework and capital accumulation |
| 1962 | Polanyi [62] | Individual | scientific discoveries as puzzle solving |

TABLE S1: List of key references on exploration and exploitation

## S3. ROBUSTNESS CHECK

### S3.1. Different timing of hot streaks

Does the observed relationship between exploration, exploitation and hot streak depend on when hot streak occurs within a career? To test this, we split artists, directors and scientists into early and late hot streak, and compare the distribution $P(H)$ for work produced before and during a hot streak for individuals with different timing of hot streak. We find that $P(H)$ during hot streak is significantly smaller than before for both cases, suggesting that our findings are robust to different timing of hot streaks (Fig. S15)

### S3.2. Different levels of impact

Does the observed exploration-exploitation transition apply to individuals with different level of impact? To test this, we identify for each individual the highest impact work within a career [1, 16], and group individuals by their highest impact into high- and low-level. We compare the distribution $P(H)$ for work produced before and during hot streak, finding that $P(H)$ during hot streak is significantly smaller than before for individuals with different level of impact across three domains (Fig. S16). We also calculate $(H)$ for work produced before and during hot streak, and compare to the entropy distribution $P((H))$ of the null model, finding again the same conclusion (Fig. S17).

### S3.3. Different disciplines

To test if our results apply to scientists from different disciplines, we identify for each scientist her discipline with subject categories provided by WoS, and group the subjects into six general disciplines: Physical Science, Biology, Medicine, Environmental Science, Chemistry and Engineering [116]. For each scientist, we count the number of papers published in each of the subject, and consider the one with the most publications as her home discipline. We repeat the analysis on $(H)$ for scientists from each of the six disciplines, and compare to the distribution $P((H))$ for 1000 realizations of the null model, finding our results remains the same (Fig. S18A-L) We also measure $P(H)$ for papers before and during hot streak in real careers, and reach the same conclusion (Fig. S18M–R).

### S3.4. Individual fixed effect

Individuals may have different baseline exploration rates. Do individuals with overall low career entropy show similar exploration-exploitation dynamics? To test this, we identify the typical level of exploration in each career, defined by the percentage of the unique number of styles/topics $n_i$ over all works one produced ($N$), denoted as $n_i/N$. We group individuals into high and low exploration rates, and compare the distribution $P(H)$ for works produced before and after during hot streak. We find that the exploration-exploitation transition occurs for individuals with different levels of entropy across the three domains (Fig. S19). We also calculate $(H)$ in real careers and compare to the entropy distribution $P((H))$ of the null model, and find that our results remain the same across three domains (Fig. S20). We further compare the level of entropy change by measuring the distribution $P(H)$ of real careers over that of the null model, denoted as $R(H) = P(H)/P_r(H)$ for works produced before and during hot streak (Fig. S21). We find that individuals tend to deviate from their typical strategy around the beginning of a hot streak: individuals who tend to exploit on average become more exploratory before a hot streak begins (Fig. S21A–C), whereas individuals who tend to explore become particularly focused during hot streak (Fig. S21D–F). This is also validated by the difference between $(H)$ and that of the null model (Fig. S20). For example, the difference between $(H)$ and null model before hot streak for low-diversity artists is more pronounced than that of the high-diversity artists. Similarly, the difference between $(H)$ and null model during hot streak for high-entropy directors is more pronounced than that of the low-entropy directors.

### S3.5. Regression analysis

In this section, we systematically calculate the correlation between diversity and the beginning of a hot streak by controlling individual-specific characters using OLS regression. We first study the entropy change when individuals started to explore before a hot streak begins, compared to their earlier career of the same length, after controlling for impact, career stage, and other individual characteristics:

$$H = a_0 + a_1 \times Stage_{HS} + a_2 \times N_i + a_3 \times career\ stage + a_4 \times t_i/N \\ + a_5 \times career\ age + a_6 \times (impact) + a_7 \times \log N \tag{S1}$$

where $Stage_{HS} = 1$ if the work is produced before a hot streak, and 0 in earlier careers; $N_i$ is the unique number of styles and topics within a career; $N$ is the overall productivity; *career stage* is the relative timing of the work overall all works produced in a career; *career age* is the years since the first work, *(impact)* captures the average impact of all works. We further compare the regression coefficient $a_1$ of real careers to that of a null model, where we randomly selected a work as the beginning of a hot streak. We then repeat the analysis for diversity change after a hot streak begins, and assign $Stage_{HS} = 1$ if the work is produced during a hot streak, and 0 before it happens. Compared to the null model where $a_1$ is rather flat, we find that $a_1$ for real careers across three domains show obvious decreasing trend from before to after a hot streak begins (Fig. S22). When individuals start to explore before a hot streak, $a_1$ is larger than the null model would expect, and $a_1$ for directors and scientists are systematically larger than 0, indicating higher diversity before a hot streak begins. While after the beginning of a hot streak, $a_1$ becomes systematically smaller than null model, indicating a drop of diversity during exploitation. Overall, our conclusions remain the same after controlling for individual specific properties.

### S3.6. Scientists with two hot streaks

Do individuals with two hot streaks experience the exploration-exploitation transition in both cases? Given that the careers with multiple hot streaks are uncommon, here we only focus on scientists who have two hot streaks. We calculate $P(H)$ for papers around each hot streak (Fig. S23), finding $P(H)$ before a hot streak is systematically larger than $P(H)$ during hot streak for both hot streaks.

### S3.7. Scientists with hot streaks at the beginning of their career

By definition, the exploration-exploitation transition can only be measured for individuals who have produced a number of works before their hot streak begins. What about the individuals whose hot streak occurs at the beginning of their careers? To test this, we focus on scientific careers with hot streak at the beginning and compare their entropy dynamics to that of the null model (Fig. S24a), or their cohorts who do not have hot streaks at the beginning (Fig. S24b). We find that in both cases, although we could not observe the behavior before their first record, individuals with hot streaks at the beginning have systematically smaller entropy, consistent with exploitation behavior during hot streak.

### S3.8. Alternative community detection method

In the main text we report the topics detected using the same community detection methods in Zeng *et al.* [18]. To test if our results are robust to different community detection algorithms, here we use Infomap [117] to detect the community structure and repeat our analysis of $P(\langle H \rangle)$, finding our conclusions remain the same (Fig. S25).

### S3.9. Different time window

In the main text we measure topic entropy for works produced during the hot streak and the prior period of the same length before. To test if our results are roust to entropy measured by different time window, we calculate the entropy for works produced within 5 years before and after the onset of hot streak (Fig. S26), and the same number of works produced before and after the onset of hot streak (Fig. S27), finding that our conclusions remains the same.

### S3.10. Alternative diversity index

To test if our results are robust to other diversity measures, here we use Simpson diversity ($D = 1 - \Sigma_i p_i^2$, where $p_i$ is the probability of topic $i$), and normalize it by the maximum value. We repeat the measurement for $P(\langle Simpson \rangle)$, finding again that $\langle Simpson \rangle$ for works before a hot streak is systematically larger than the null model across the three domains (Fig. S28A–C) Similarity, $\langle Simpson \rangle$ for works after a hot streak begins is again systematically smaller than expected (Fig. S28D–F). Together, the uncovered exploration and exploitation transition remains the same for different diversity measures.

### S3.11. Episodes of exploration and exploitation

To systematically understand the correlation between exploration, exploitation and the onset of career hot streaks, we define for each individual episodes of exploration and ex-ploitation within a career by calculating the style or topic entropy in a sliding window of two years for artists and scientists, and five films for directors. We calculate the probability to initiate a hot streak at the end of an exploration episode ($P_\downarrow$), at the beginning of an exploitation episode ($P_\uparrow$), at the transition from exploration to exploitation ($P_{\downarrow\uparrow}$), and at the transition from exploitation to exploitation ($P_{\uparrow\downarrow}$), We further compare their relative change to the baseline probability $P_r$, defined as the average probability for the beginning of an episode to coincide with a hot streak among all episodes in a domain (0.040 for artists, 0.042 for directors, and 0.073 for scientists, respectively). In the main text (Fig. 3S-U), we report the relative change in the probability ($P = P_{\downarrow,\uparrow,\downarrow\uparrow,\uparrow\downarrow}/P_r - 1$).

### S3.12. Different numbers of centroids

We test in this section whether our results are robust to different number of clusters for artists and directors. We retrain the k-means clustering with 20 and 40 centroids and repeat the analysis of $P(\langle H \rangle)$, finding the results are robust with different numbers of centroids (Fig. S29).

### S3.13. Papers without references

We find that some of the papers do not have references in WoS, which lack sufficient information for us to identify their topics from the co-citing network. We ignore papers without references in scientific careers in the main text. To test whether our results are robust if we include those papers, we include the impact of papers without references when we measure hot streak, use the same community detection methods to identify topics, and assign papers without references to a new topic. We then calculate the entropy distribution $P(\langle H \rangle)$ for 10 works produced before and during hot streak begins and compare to the null model, finding again that papers produced before a hot streak have higher entropy than expected, while papers during a hot streak have significantly low entropy than expected (Fig. S30).

### S3.14. Different style measurements

In this section, we test if the results are robust if we use different style measurements for artists and directors. We directly use style labels for artworks and genres labels for films to measure entropy distribution. We use fine-tuned VGGNet to predict the style of each image. The model output is a 29-dimensional vector with probability to each style, and the image is assigned the style with the highest probability. We then calculate the entropy distribution $P((H))$ for works produced within 5 years before and during hot streak, finding our conclusion remains the same (Fig. S31). We also use genre labels in IMDB to approximate the style of each film. We focus on the genres for 5 works produced before and after a hot streak begins. If a film has multiple genres, we include all of them in the genre list. We again observe the transition from exploration to exploitation if we measure film style with genres (Fig. S32), suggesting that our analysis is robust under different style definitions.

## S4. SCIENTIFIC TEAMS

### S4.1. Regression analysis on team size

In the main text, we compare the team size during exploration and exploitation for all scientists in our dataset. To ensure that our results are not affected any temporal trend or population differences, here we perform a OLS regression to study the correlation between team size and the beginning of a hot streak. We first measure the change of team size when scientists started to explore before a hot streak begins compared to their earlier career of the same length, after controlling for year, career stage and research fields:

$$\log \textit{team size} = a_0 + a_1 \times \textit{Stage}_{HS} + a_2 \times \textit{year} + a_3 \times \textit{career stage} \\ + a_4 \times \textit{career age} + a_5 \times \textit{field} \quad \text{(S2)}$$

where log *team size* is the logrithmic of team size for a paper, as the distribution of $P(\textit{team size})$ is fat-tailed in general (Fig. S33). $\textit{Stage}_{HS} = 1$ if the paper is produced before a hot streak (within 10 papers), and 0 if the paper is produced in one's earlier career; *career stage* captures the relative timing of the paper overall all papers produced in a career; *career age* measures the years since the first paper; and field is the subject category in WoS which the paper belongs to. We further compare the regression coefficient $a_1$ of real careers to that of a null model, where we randomly selected a paper as the beginning of a hot streak. We then repeat the analysis for team size after a hot streak begins, and assign $Stage_{HS}$ = 1 if the work is produced during a hot streak (within 10 papers), and 0 otherwise.

Compared to the null model where $a_1$ is rather flat, we find that $a_1$ for real careers show increasing trends from before to after the onset of a hot streak (Fig. S34). $a_1$ for the exploration phase is smaller than the null model would expect, and is systematically smaller than 0, indicating that scientists explore with smaller teams before a hot streak begins. While after the beginning of a hot streak, $a_1$ becomes systematically larger than null model, indicating that scientists work with larger teams during exploitation. Overall, our results are robust after controlling for temporal and disciplinary differences.

### S4.2. Different disciplines

We further split scientists into six major domains following S6.4, and run the OLS regressions for scientists in each domain separately (Fig. S35). Consistent with prior results, we find that the team size across six domains shows significant increase after a hot streak begins. $a_1$ for the team size before hot streak is smaller in real careers than that of the null model, and scientists from physical science, environmental science and engineering have more pronounced effects.

### S4.3. Team composition

The large team during hot streak may be not simply expanded from past collaborators. Rather, scientists may work with a new group of collaborators following the onset of hot streak. To test this, we calculate the dynamics for the number of common authors $A_{shared}$ for 5 papers before and after a given position $t$, divided by the total number of unique authors $A$ for the 10 consecutive papers (Fig. S36A). Here we focus on the individuals with similar hot streak duration ($L$ = 10 ± 2) and align their careers by the timing of the hot streak. $A_{shared}/A$ at the beginning of a hot streak is significantly smaller than the null model, suggesting that collaborators after $t_1$ are less likely to overlap with collaborators before. We further validate this result by measuring the rate of new co-authors for papers. Specifically, we calculate the number of new co-authors $A_{new}$ within a sliding window of 5 papers, divided by the number of unique authors $A$ during the same time. We find that hot streak begins with the highest rate of new collaborators as $A_{new}/A$ peaks at $t_\uparrow$, and is significantly higher than the null model (Fig. S36B). Together, Fig. S36 suggests that instead of expanding their collaborators when a hot streak begins, scientists are more likely to work with a different group of collaborators during their hot streak.

## S5. CHARACTERISTICS OF EXPLOITED TOPICS DURING HOT STREAK

In this section, we probe the connections between phases of exploration and exploitation and examine properties of the topics that are explored before a hot streak begins, ranging from their recency to citation impact to popularity (Fig. S37), asking which ones tend to be chosen for subsequent exploitation.

### S5.1. Recency

Prior research shows that topic evolution along a scientific career is characterized by recency [64], implying that scientists should exploit during hot streak the latest topic they explored. We test this hypothesis by calculating the probability of the exploited topics to be the most recent before a hot streak begins (Fig. S37C, left). Among exploited topics that are studied before, around 32.7% of them are the most recent ones, lower than the null model would expect (33.6%, Chi-square test, p-value=0.0019). Thus, as individuals may learn from exploration to deliberately find a direction worth going deep, the transition from exploration to exploitation when a hot streak begins is not simply due to perceiving a new direction by chance and reaping its benefits.

### S5.2. Popularity in a career

Prior study also shows that scientists have their core research topic that are repeatedly investigated [64], suggesting the exploited topic may be popular within a career. To test this, we measure the popularity of a topic by the number of paper published before the onset of hot streak, and calculate the probability for the topic exploited to be the most popular. We find that the probability to select the most popular topic is lower than expected (Chi-square test, p-value=$6.12 \times 10^{-40}$), suggesting that scientists are less likely to continue their past focal topic during hot streak.

### S5.3. Impact

Is the exploited topic the highest cited among all topics explored before? We compare the impact for topics before a hot streak by measuring the average paper citation till the onset of hot streak, and categorize explored topics into low, middle and high impact. We calculate the probability for the exploited topic to fall into each category, and find that the topics studied during exploitation are less likely to be the high-impact topic (Fig. S37D).

### S5.4. Popularity in embedding space

Exploration may increase the likelihood for individuals to stumble upon a hot topic and reap its benefits during hot streak, prompting us to test if the exploited topic is popular at that time. We project the 64-dimensional embeddings learned from co-citing network among all papers in the dataset onto two dimensions using PCA. We calculate the density on these dimensions by measuring the volume of papers at the time in the embedding space. The popularity of a topic is defined as the average density for papers belonging to the topic when they were published. We find that the topics studied during exploitation are less likely to reside in either high- or low-density regions but rather somewhere in the middle (Fig. S37E). We also calculate the momentum of each topic during the exploration phase, which traces the rate of increase in popularity within five years before the onset of hot streak, finding again that the topic exploited is not among the fastest growing (Fig. S37E, inset)

### S5.5. Predicting topic exploited

We utilize the topic properties discussed above together with the team size to predict which topic a scientist will choose to exploit during their hot streak. Specifically, we formulate a binary classification problem: For each scientist, given each topic she explored before hot streak, we predict whether it will be exploited during the hot streak based on its recency, impact, popularity and team size. Here we focus on careers whose exploited topics were among those studied before (around 80% among all scientists in the dataset). We calculate the topic recency, impact, popularity in a career and in the embedding space following th same procedure above. The team size of each topic is calculated as the average log team size for papers belonging to the topic that were published before hot streak.

We randomly sample 80% of topic records as training set and the rest 20% as the test set, and use the random forest model with 500 trees to train the classifier. The model accuracy is 0.89 and AUC is 0.83 (Fig. S37F). We also test the effect of imbalanced positive (exploited) and negative (not exploited) sample size, and down-sample the negative cases to the same amount of positive ones, finding the AUC remain above 0.8. We further compare the model accuracy to two baselines: 1) the accuracy for the topic from the last paper before hot streak is 0.64; 2) the accuracy for a randomly selected topic before hot streak is 0.43, finding our prediction model significantly outperforms the two baselines.

## S6. TESTING ALTERNATIVE HYPOTHESIS

To explore alternative explanations for the onset of hot streaks, we test several hypotheses in this section, each capturing potential individual or institutional factors that may affect career progression and success:

**Multiple publications**. Innovators may stumble upon a groundbreaking idea, which manifests itself in the forms of multiple artworks, films, or publications. Hence from an evolutionary perspective, hot streak may correspond to the duration for the temporary competitive advantage to dissipate. We test this hypothesis by measuring the relative order of three highest impact papers during hot streak. Multiple publications hypothesis predicts that the highest impact paper should be more likely to occur before the second highest. By contrast, we find that there is an equal probability for the highest-impact work to appear before or after the second highest (Fig. S38A).

**New research direction.** Some research topics are more impactful than others, and hot streaks may be simply driven by switching to a new research direction, which affects an individual's overall achievements. We test this hypothesis by measuring the probability of observing a new topic when hot streak begins, and compare it with that of the randomized careers (Fig. S38B). However, we find no detectable difference between data and the null model (Chi-square test, p-value =0.23).

**Collaboration with high-impact scientists**. Teams are increasingly responsible for producing high-impact work, suggesting hot streak may reflect fruitful, repeated collaborations with other high-impact individuals. To test this, we quantify the reputation of one's coauthors using h-index, and measure the h-index distribution for the most prestigious co-author of each paper published during hot streak. We find that the h-index for the most prestigious coauthor is largely indistinguishable from the null model (Fig. S38C).

**Changing institutions**. Scientists moving to a new intellectual environment may be exposed to new sets of ideas or opportunities, which may increase their likelihood to produce high-impact work . To investigate the relationship between changing institutions and hot streak, we trace the physical mobility of scientists through their affiliations recorded in publication records [118], and calculate the probability of changing institutions during their hot streak. We find that hot streaks are less likely to be associated with affiliation change than expected (Chi-square test, p-value = $1.8 \times 10^{-41}$, Fig. S38D).

**Research support**. A new grant may help accelerate a scientist's research progress and offer opportunities to produce high-impact papers. To test whether hot streak can be explained by new grants, we linked the careers of scientists in our dataset with a funding database that captures 3.7 million funded projects across more than 250 funding agencies worldwide (https://www.dimensions.ai/) by the same last name, first initial and affiliation. We measure the number of new grants around the onset of a hot streak, and find no detectable differences compared with the null model (Fig. S39E, Chi-square test, p-value = 0.36). We further calculate the amount of funding one received around the beginning of a hot streak, finding again a lack of difference between data and the null model (Fig. S38E, KS test, p-value = 0.67).

**Future hot topic**. Although scientists are less likely to work on the most popular topic at the time, scientists may happen to work on a topic that becomes hot in the future. To test this hypothesis, we quantify the impact of each topic in the WoS data and compare the distribution of topic impact during hot streak to that of before, and find that the impact improvement for topics appears negligible relative to the overall impact change in real careers (Fig. S38F).

Overall, these results demonstrate that none of the alternative hypotheses alone can account for the onset of career hot streaks.

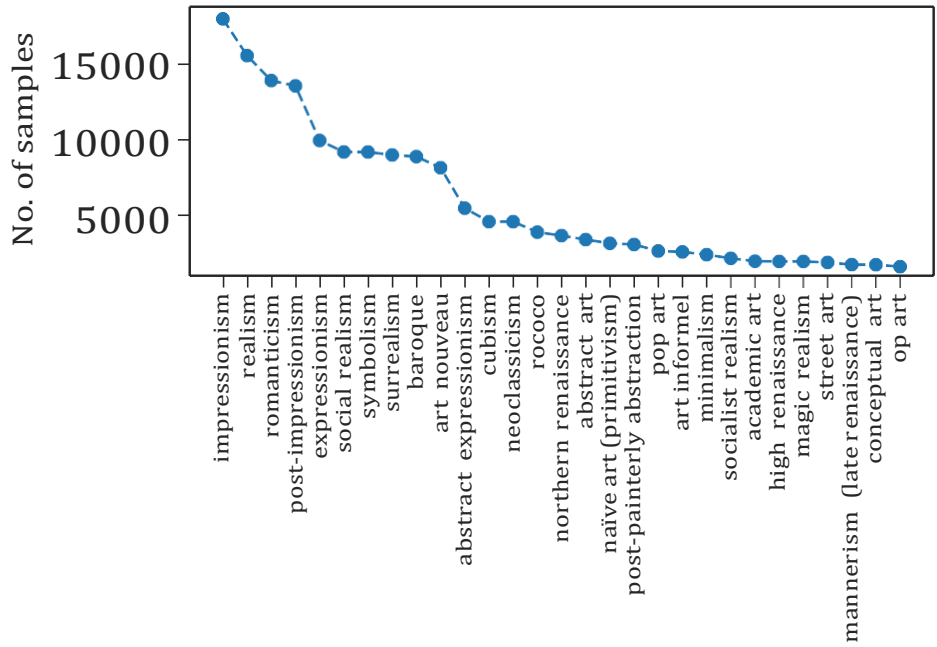

FIG. S1: The number of samples for top 20 labels in Art500k

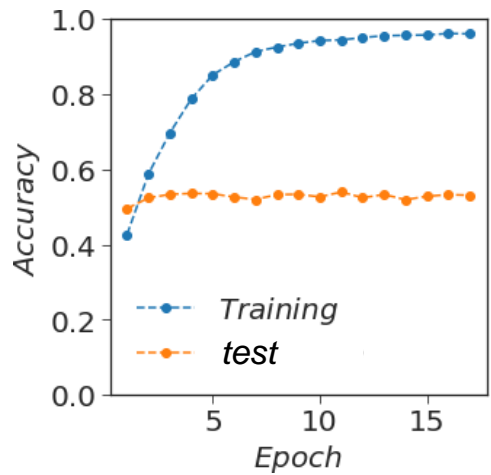

FIG. S2: The training and test accuracy of fine-tuned VGG16 to predict art style labels.

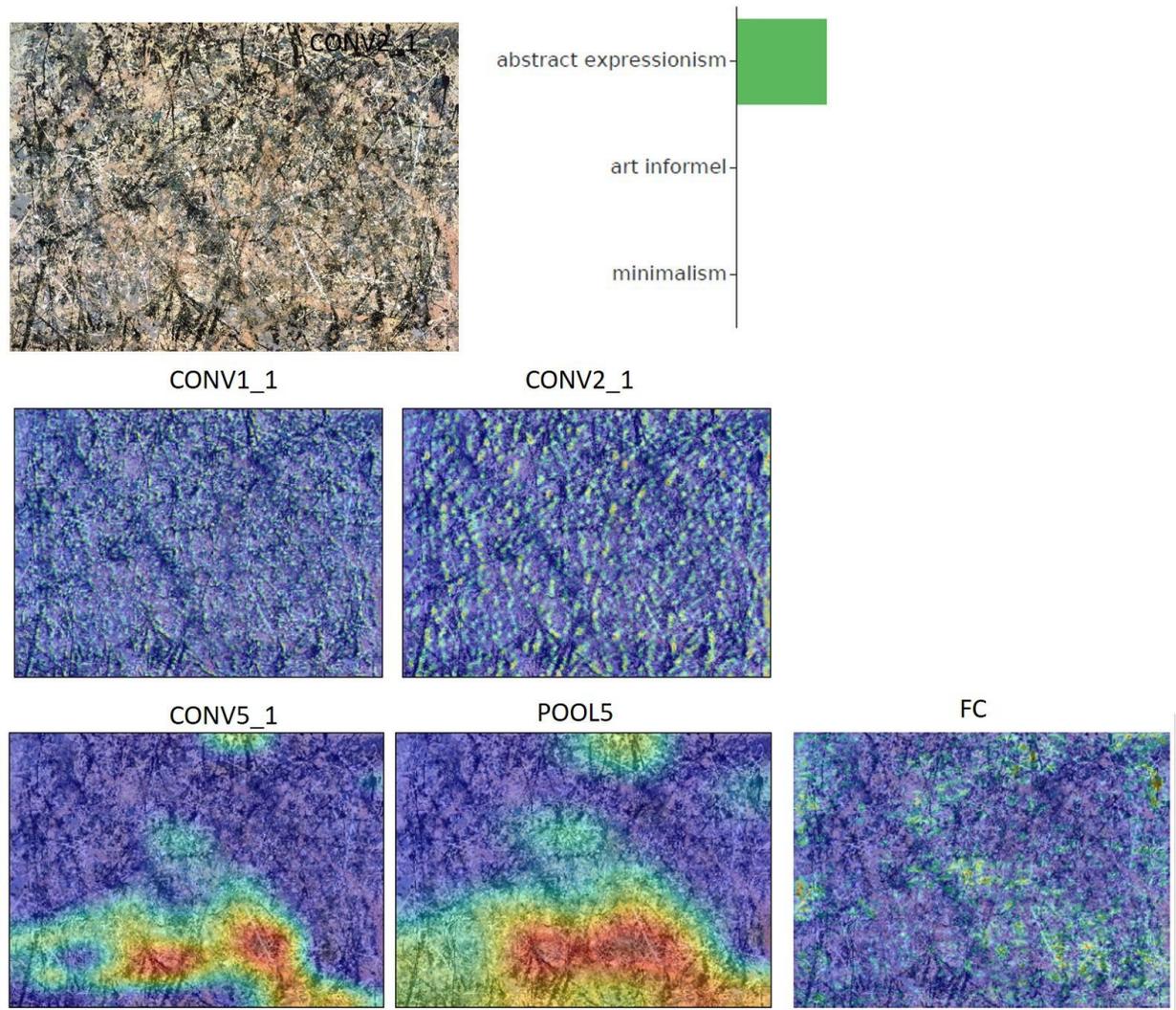

FIG. S3: A case study on model interpretation for Number 1. The top row shows the raw image and the model prediction results, with probability in top 3 styles. The middle and low panel show a random feature map from *CONV* 1_1, *CONV* 2_1, *CONV* 5_1, *POOL*5 and the saliency map with the most likely art style (*FC*).

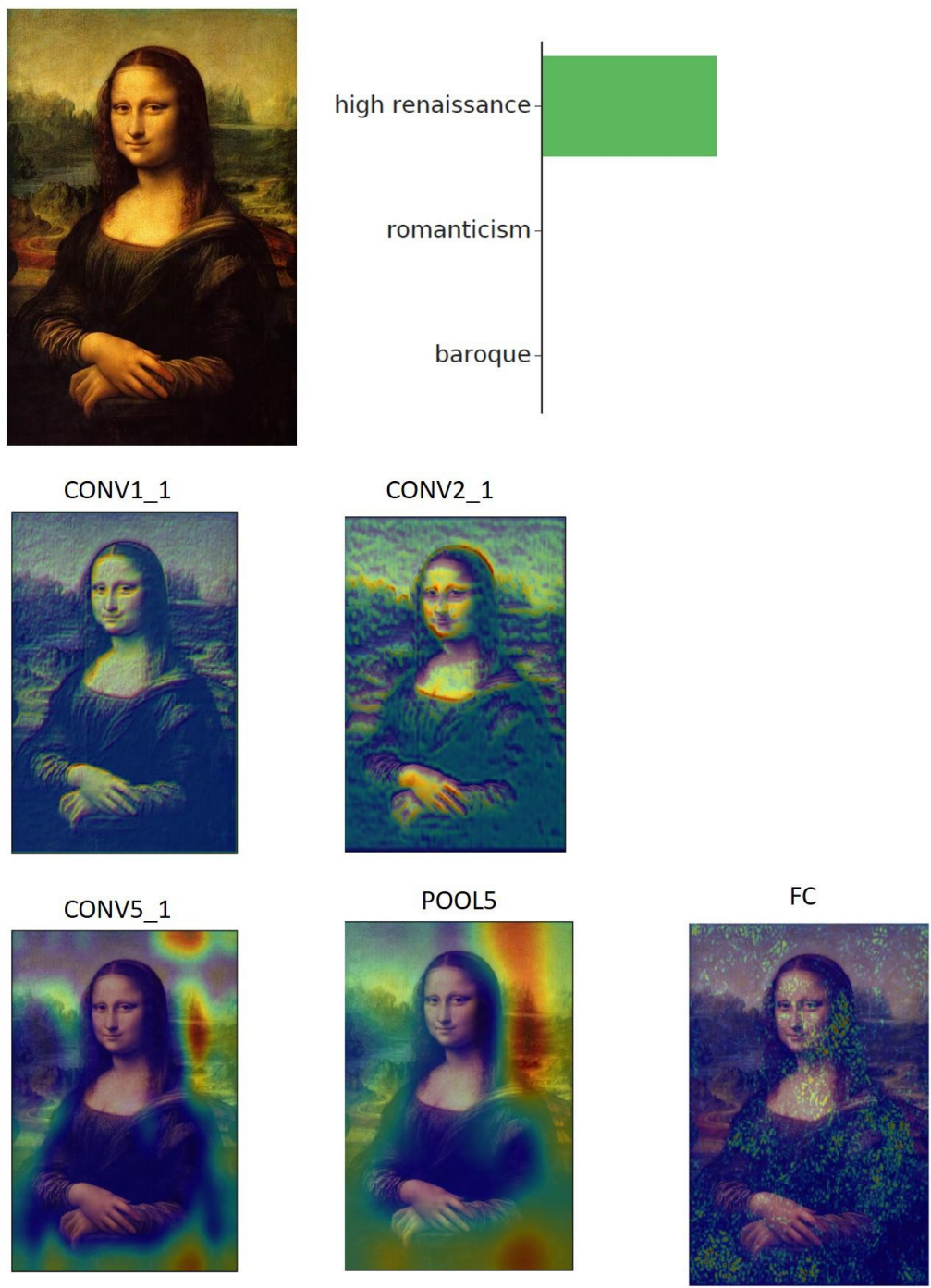

FIG. S4: A case study on model interpretation for Mona Lisa. The top row shows the raw image and the model prediction results, with probability in top 3 styles. The middle and low panel show a random feature map from *CONV 1_1*, *CONV 2_1*, *CONV 5_1*, *POOL5* and the saliency map with the most likely art style (*FC*).

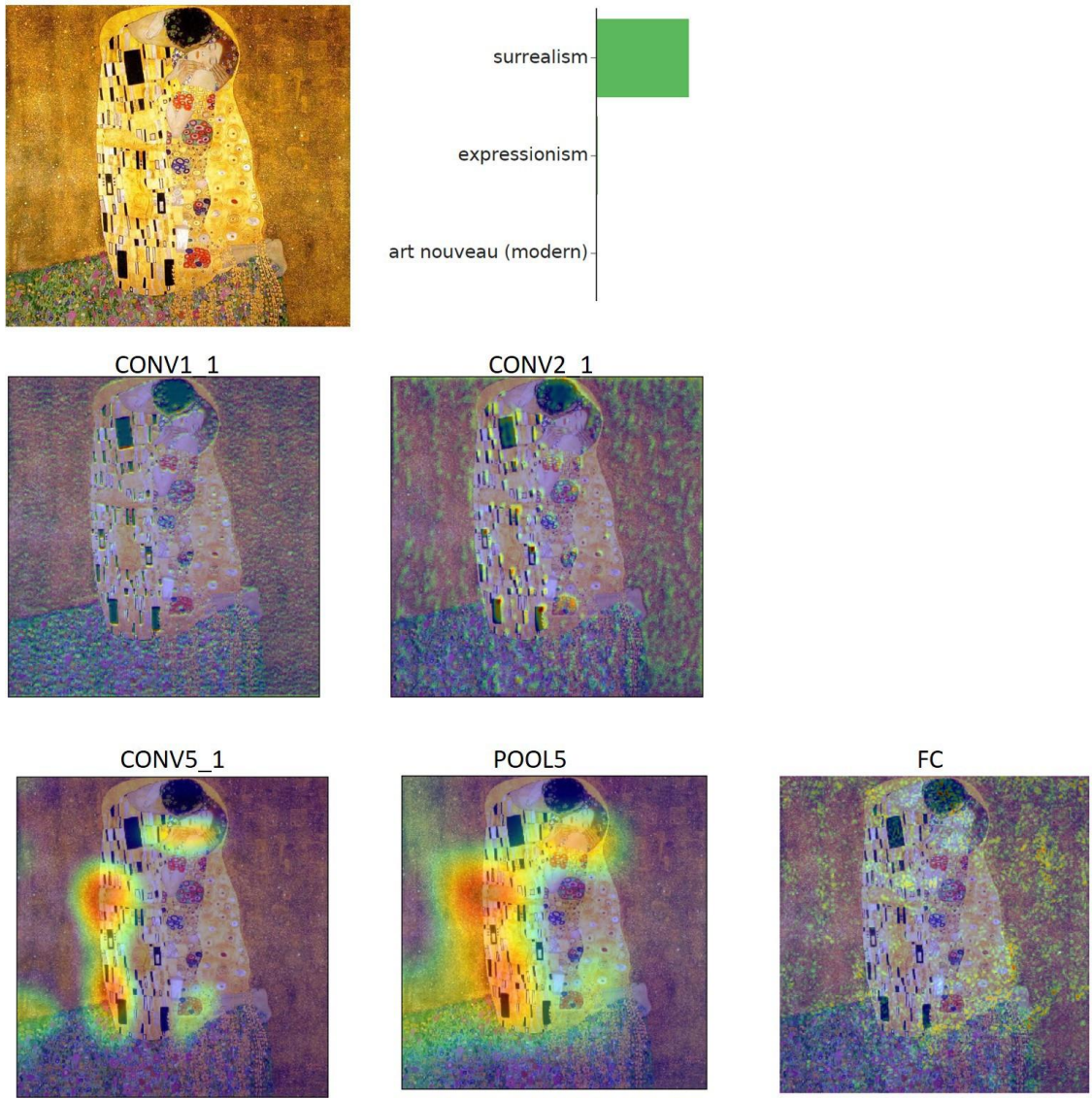

FIG. S5: A case study on model interpretation for Kiss. The top row shows the raw image and the model prediction results, with probability in top 3 styles. The middle and low panel show a random feature map from *CONV* 1_1, *CONV* 2_1, *CONV* 5_1, *POOL*5 and the saliency map with the most likely art style (*FC*).

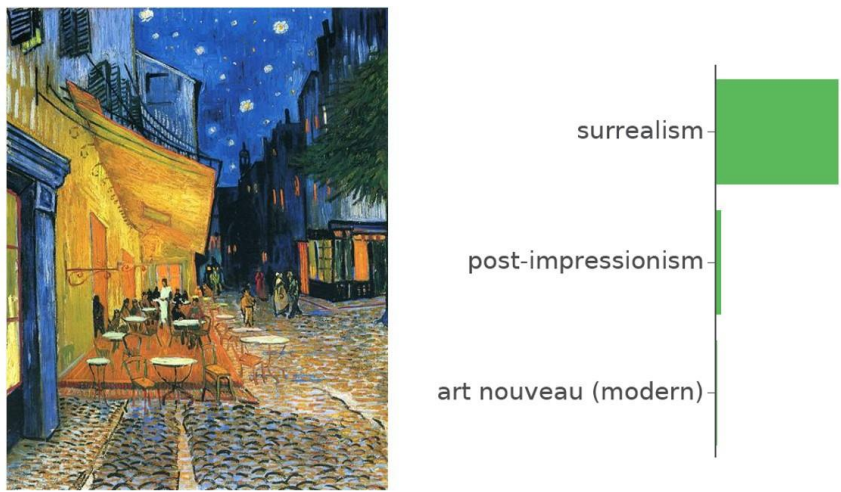

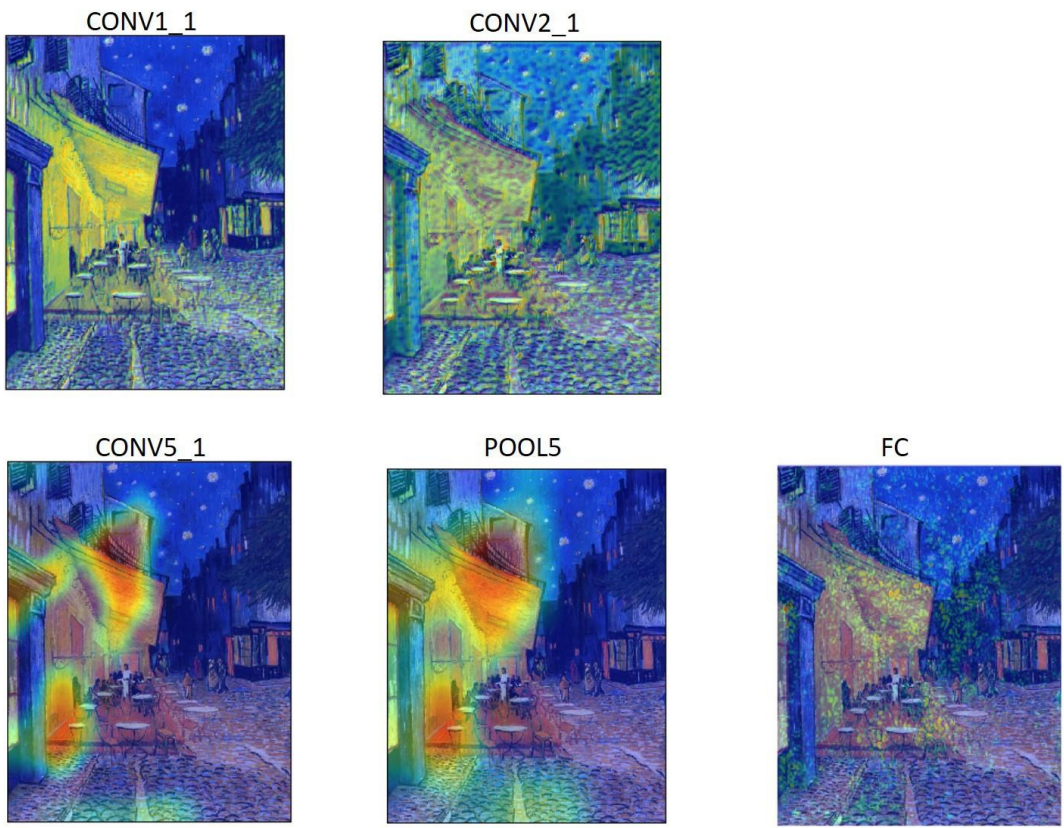

FIG. S6: A case study on model interpretation for Cafe terrace. The top row shows the raw image and the model prediction results, with probability in top 3 styles. The middle and low panel show a random feature map from $CONV\_1\_1$, $CONV\_2\_1$, $CONV\_5\_1$, $POOL5$ and the saliency map with the most likely art style ($FC$).

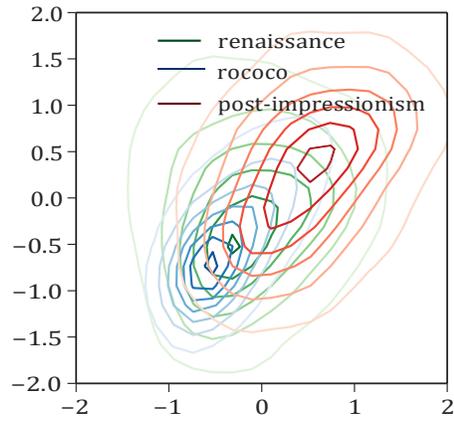

FIG. S7: The kernel density for images from renaissance, impressionism and pop art projected onto a 2D embedding space with principle component analysis

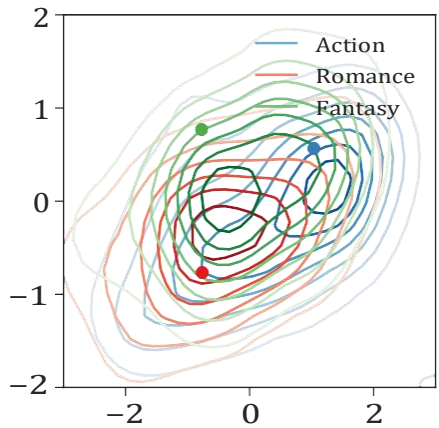

FIG. S8: The kernel density for action, romance and fantasy films projected onto a 2D embedding space with principle component analysis

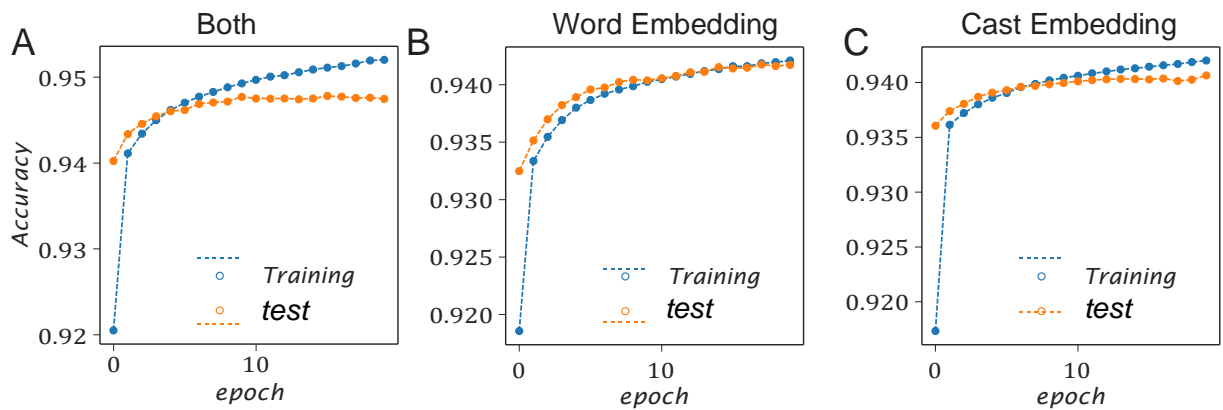

FIG. S9: The training and test accuracy for the neural network to predict film genres.

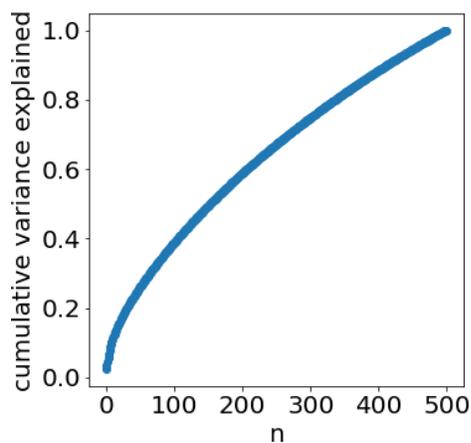

FIG. S10: The cumulative variance explained for the top 100 eigenvalues.

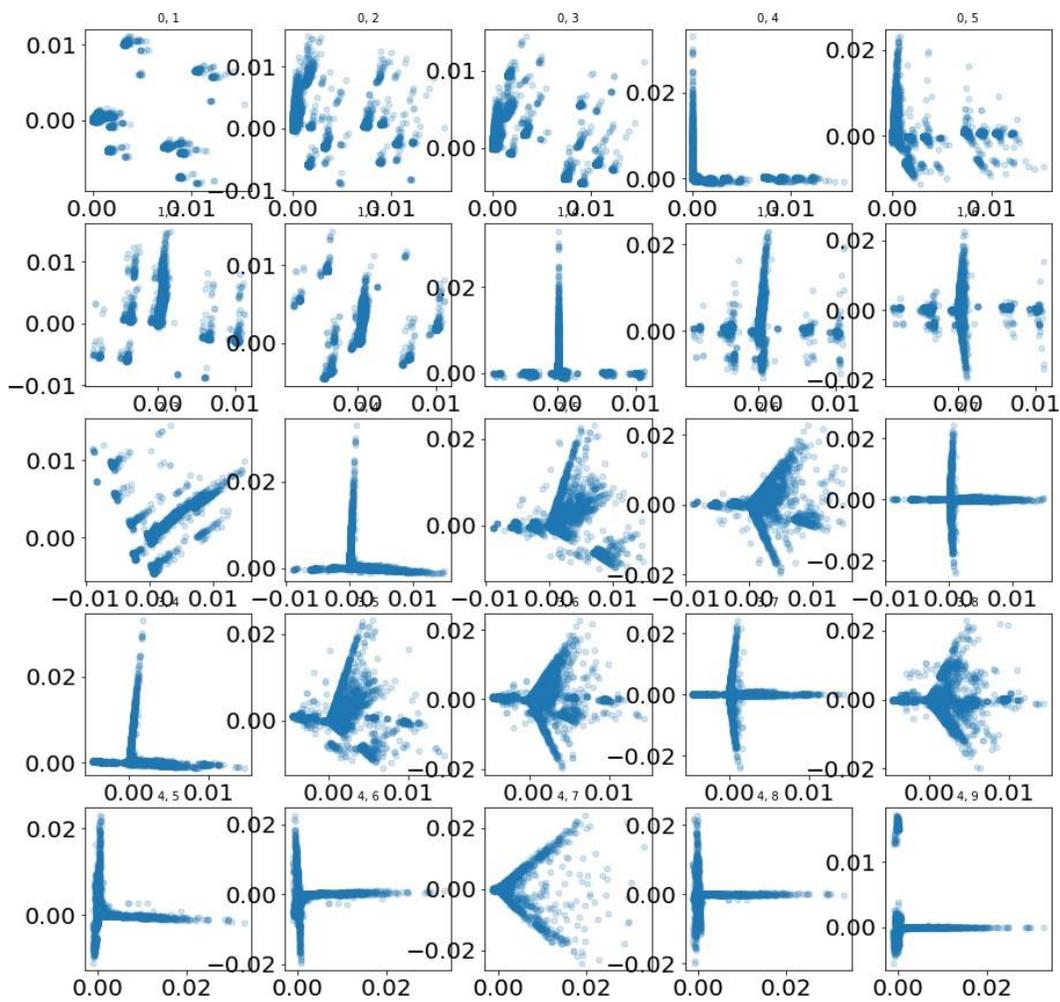

FIG. S11: The scatter plot for two columns from the u matrix, only considered top 100k samples. The title in each subplot represents the dimension we measured.

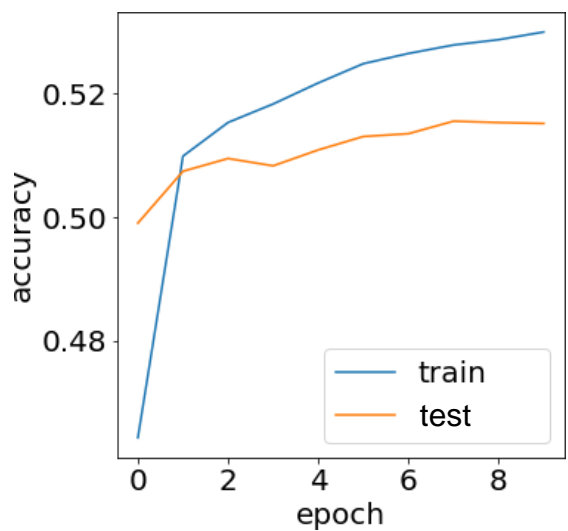

FIG. S12: The training and test accuracy for the neural network in predicting paper subject.

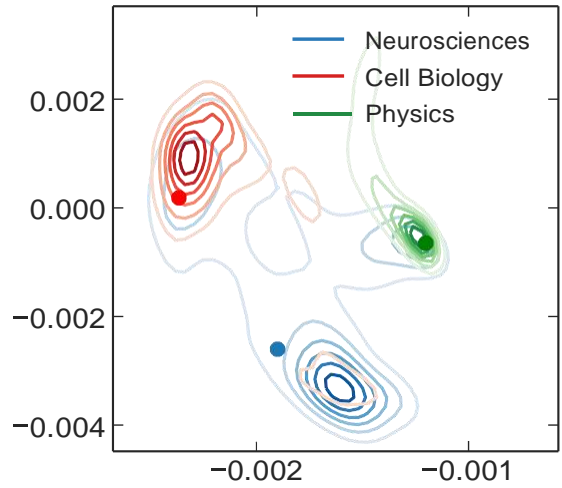

FIG. S13: The kernel density for neurosciences, cell biology and physics papers projected onto a 2D embedding space with principle component analysis.

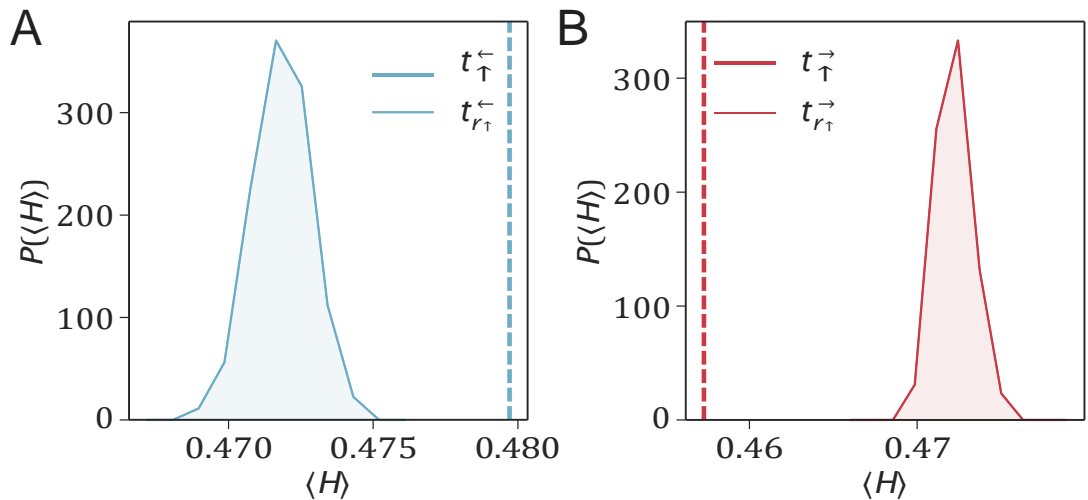

FIG. S14: The distribution of rescaled entropy $P(\langle H \rangle)$ before and during hot streak for 1000 realizations of the randomized scientific careers using topics measured from the node embedding space. $\langle H \rangle$ measured from real careers (vertical line) is significantly larger than expected before hot streak and smaller than expected during hot streak.

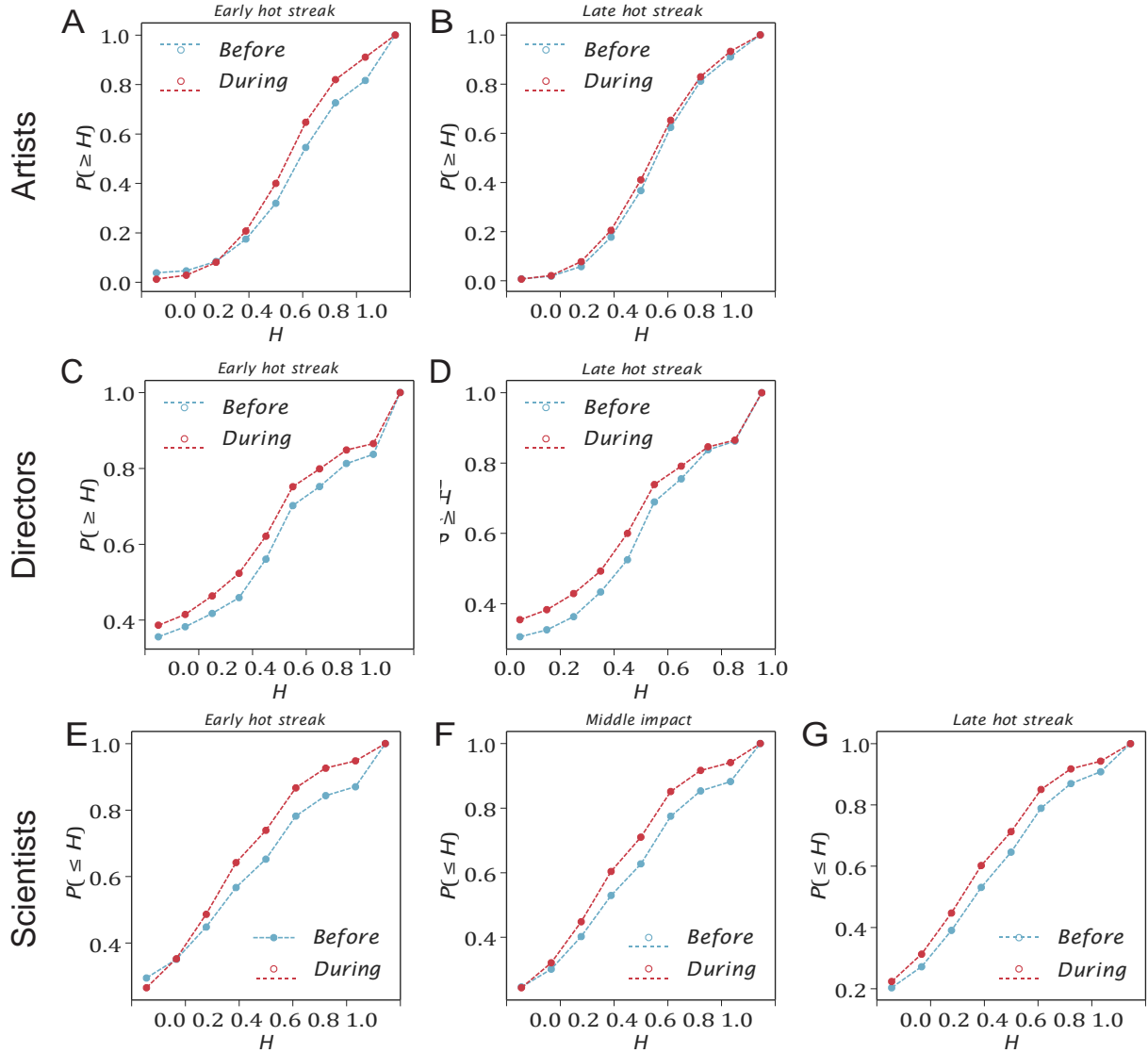

FIG. S15: The distribution of topic entropy $P(H)$ for works produced before and during hot streak for individuals with different timing of hot streak. We split artists and directors into early ($N_\uparrow/N \leq 1/2$) and late ($N_\uparrow/N \leq 1/2$) hot streak. Given that the sample size for scientists is larger than artists and directors, we split scientists into early ($N_\uparrow/N \leq 1/3$), middle ($1/3 < N_\uparrow/N \leq 2/3$) and late ($N_\uparrow/N > 2/3$) hot streak. $H$ before hot streak is consistently larger than $H$ during hot streak.

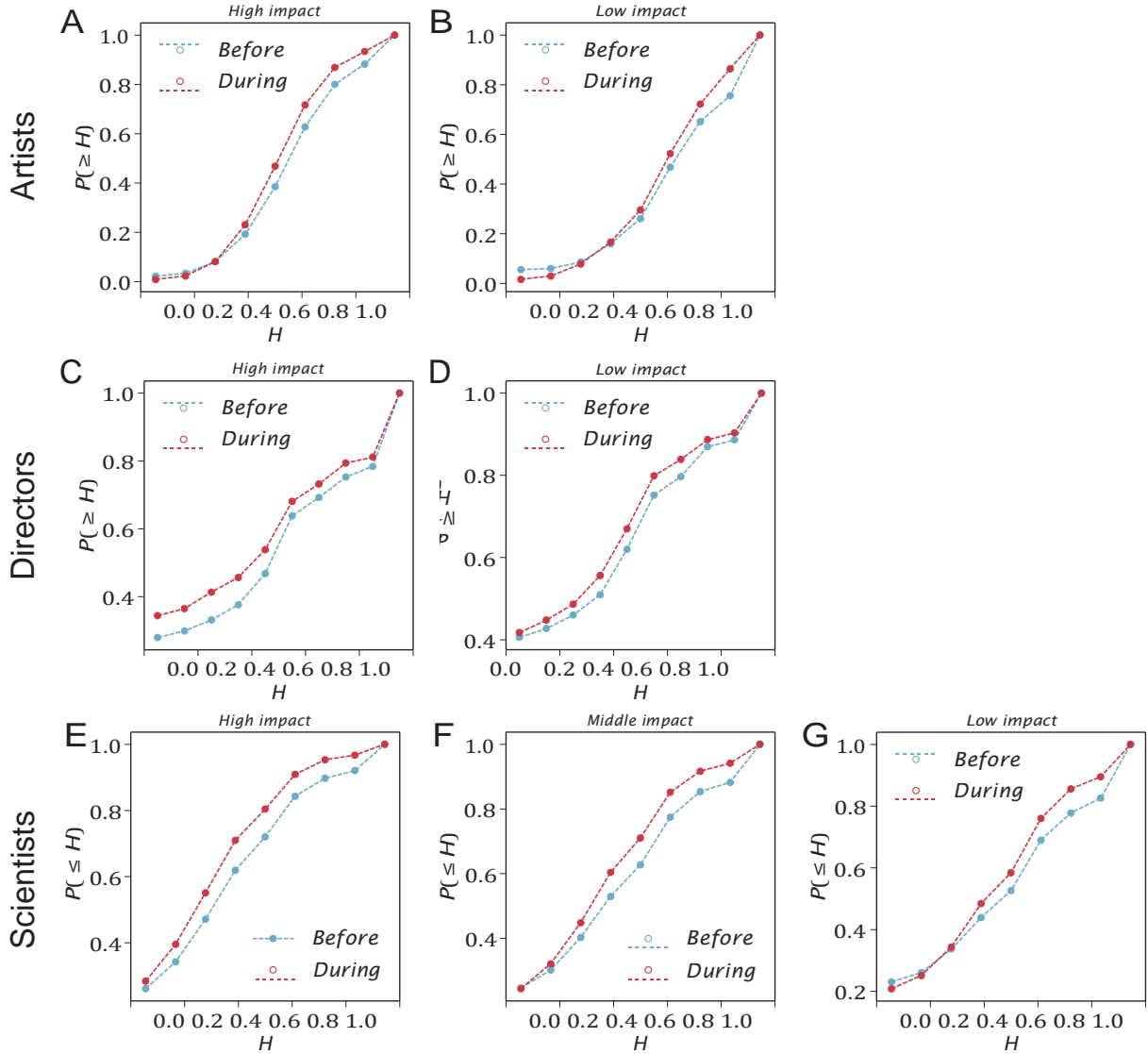

FIG. S16: The distribution of topic entropy $P(H)$ for works produced before and during hot streak for individuals with different levels of impact. We split artists and directors into high (top 1/2) and low (bottom 1/2) level of impact. Given that the sample size for scientists is larger than artists and directors, we split scientists into high (top 1/3), low (bottom 1/3), and middle (the rest 1/3) level of impact. $H$ before hot streak is consistently larger than $H$ during hot streak.

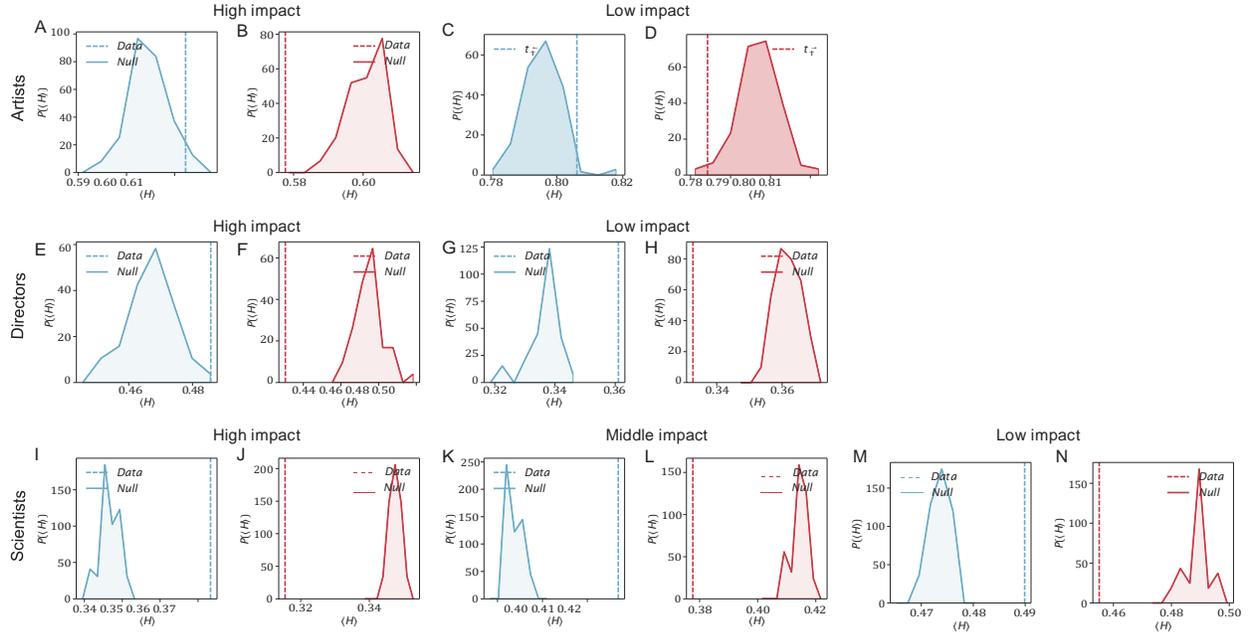

FIG. S17: The distribution of rescaled entropy $P(\langle H \rangle)$ before and during hot streak for 1000 realizations of the randomized careers before and during hot streak for individuals with different levels of impact. We split artists and directors into high (top 1/2) and low (bottom 1/2) level of impact. Given that the sample size for scientists is larger than artists and directors, we split scientists into high (top 1/3), low (bottom 1/3), and middle (the rest 1/3) level of impact. $\langle H \rangle$ measured from real careers (vertical line) is larger than expected before hot streak and smaller than expected during hot streak.

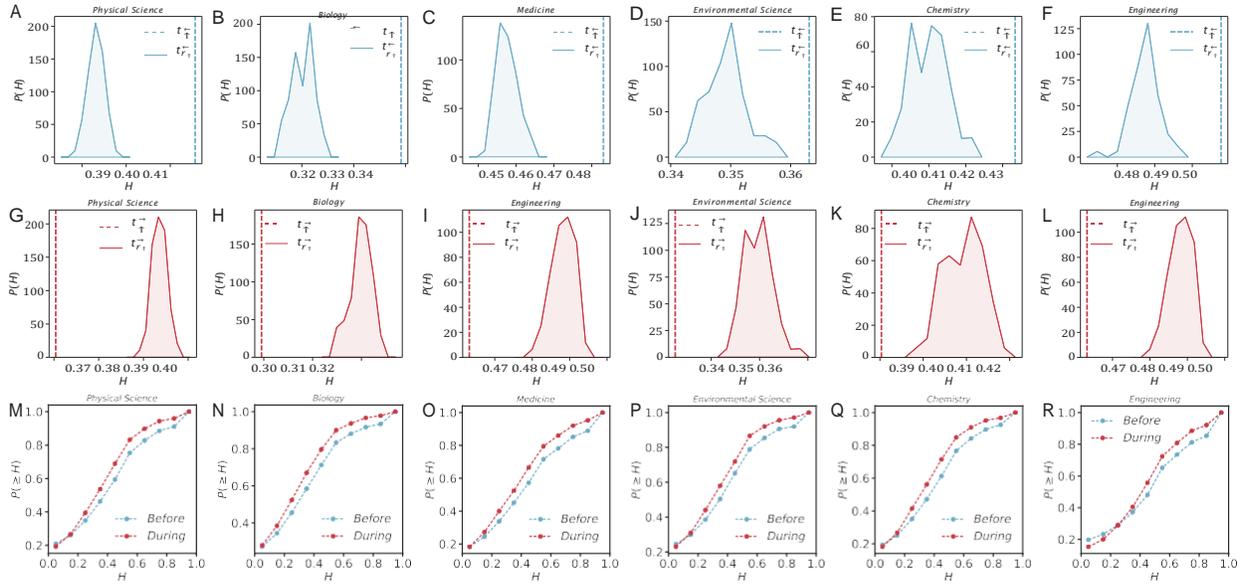

FIG. S18: (A-L) The distribution of rescaled entropy $P(\langle H\rangle)$ before and during hot streak for 1000 realizations of the randomized scientific careers before and during hot streak for scientists from six disciplines. $\langle H\rangle$ measured from real careers (vertical line) is larger than expected before hot streak and smaller than expected during hot streak. (M-R) The distribution of topic entropy $P(H)$ for works produced before and during hot streak for scientists from six disciplines. $H$ before hot streak is consistently larger than $H$ during hot streak.

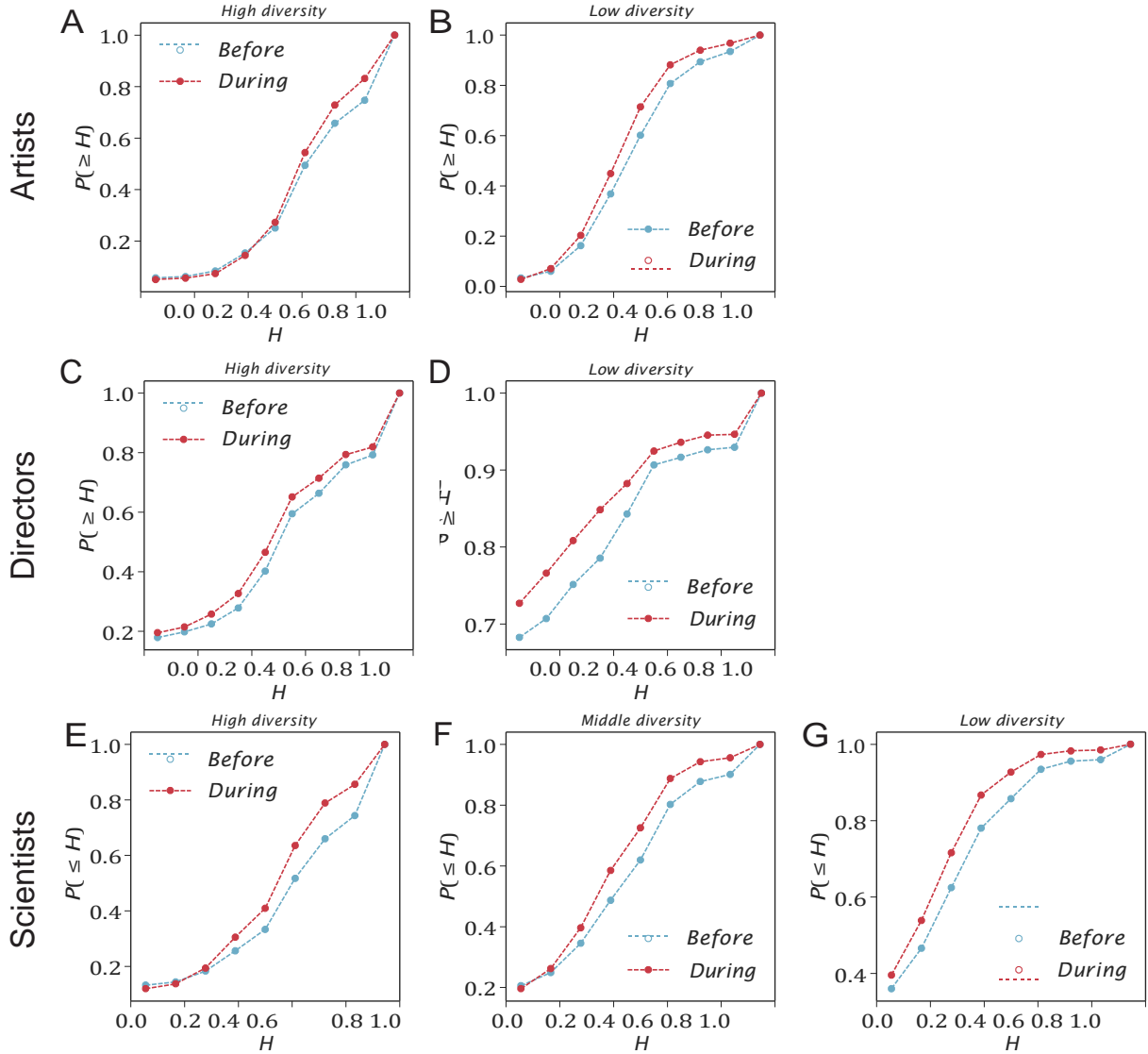

FIG. S19: The distribution of topic entropy $P(H)$ for works produced before and during hot streak for individuals with high (top 1/2) and low (bottom 1/2) baseline exploration rates $n_i/N$. Given that the sample size for scientists is larger than artists and directors, we split scientists into high (top 1/3), low (bottom 1/3), and middle (the rest 1/3) baseline exploration rates. $H$ before hot streak is consistently larger than $H$ during hot streak.

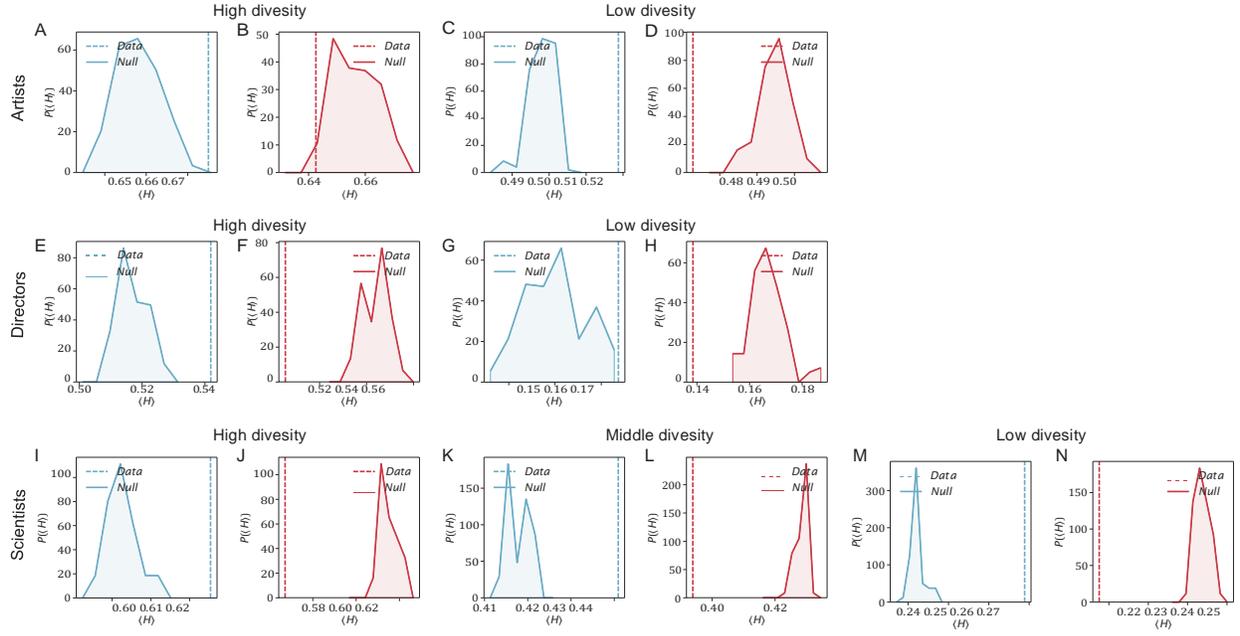

FIG. S20: The distribution of rescaled entropy $P(\langle H \rangle)$ before and during hot streak for 1000 realizations of the randomized careers before and during hot streak for individuals with high (top 1/2) and low (bottom 1/2) baseline exploration rates $n_i/N$. Given that the sample size for scientists is larger than artists and directors, we split scientists into high (top 1/3), low (bottom 1/3), and middle (the rest 1/3) baseline exploration rates. $\langle H \rangle$ measured from real careers (vertical line) is significantly larger than expected before hot streak and smaller than expected during hot streak.

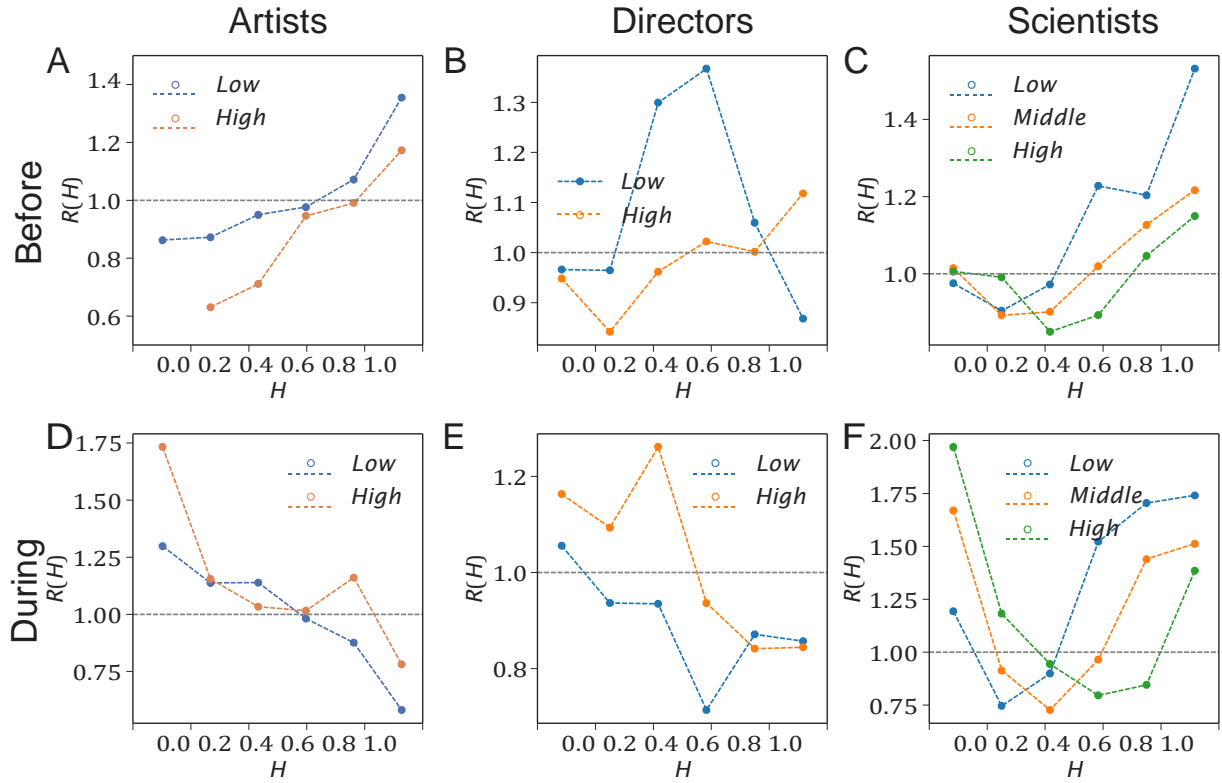

FIG. S21: We compare the relative entropy distribution between data $P(H)$ and the null model $P_r(H)$, denoted as $R(H) = P(H)/P_r(H)$ for works produced (A-C) before and (D-F) during hot streak. Individuals with low exploration rates tend to explore more than expected before hot streak, and individuals with high exploration rates tend to exploit more than expected during hot streak.

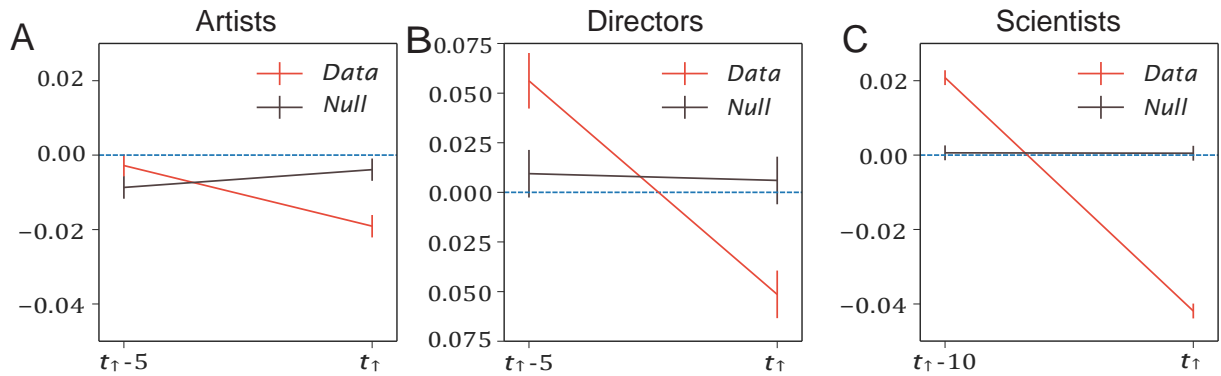

FIG. S22: Regression coefficient of entropy change on the timing of hot streak from linear regressions for data and the null model.

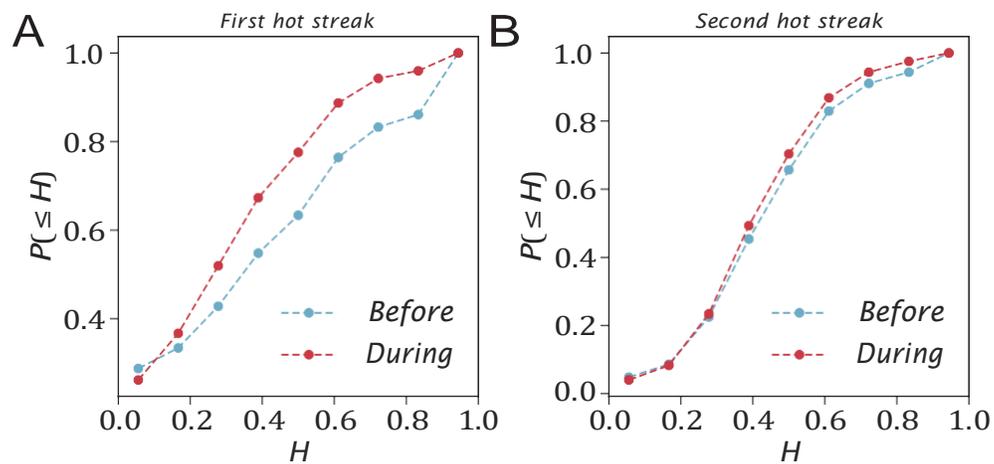

FIG. S23: The distribution of topic entropy $P(H)$ for works produced before and during the (A) first and (B) second hot streak for scientists.

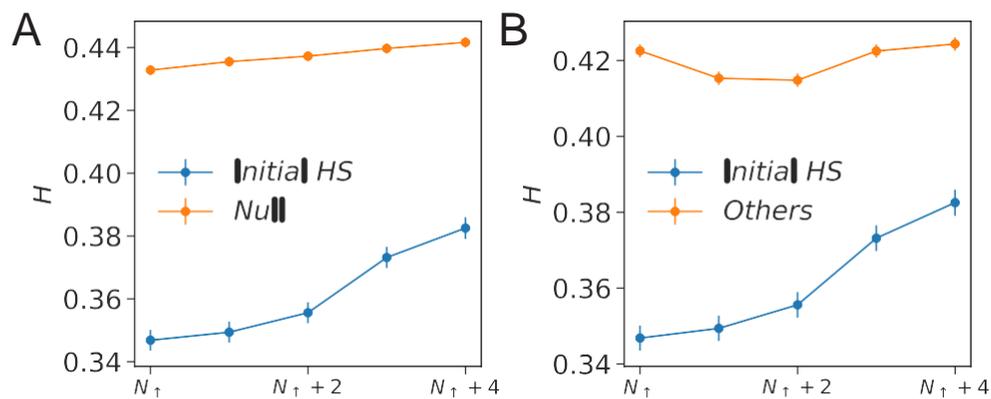

FIG. S24: The dynamics of average entropy in a sliding window of 5 papers for scientists with hot streak initiating at the beginning of their careers, compared to the entropy at the beginning of a career for (A) the null model, and (B) other scientists whose hot streak come later.

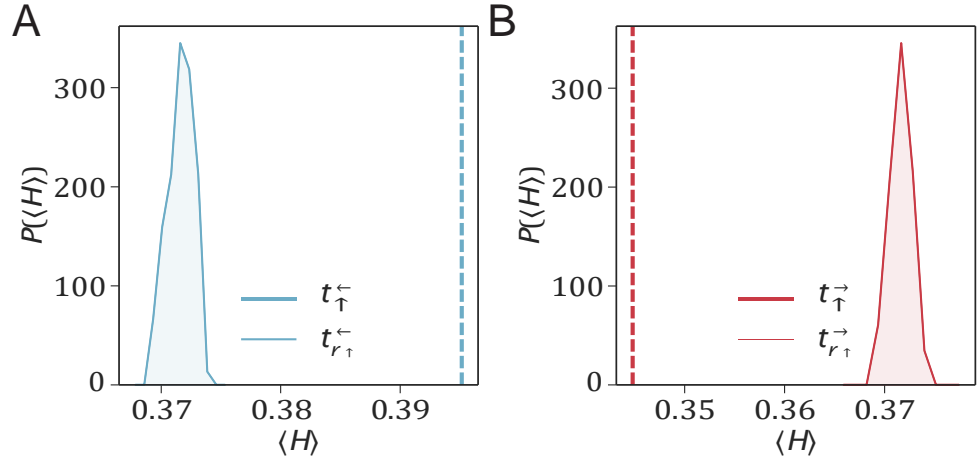

FIG. S25: The distribution of rescaled entropy $P(\langle H \rangle)$ before and during hot streak for 1000 realizations of the randomized scientific careers before and during hot streak using topics measured by Infomap. $\langle H \rangle$ measured from real careers (vertical line) is significantly larger than expected before hot streak and smaller than expected during hot streak.

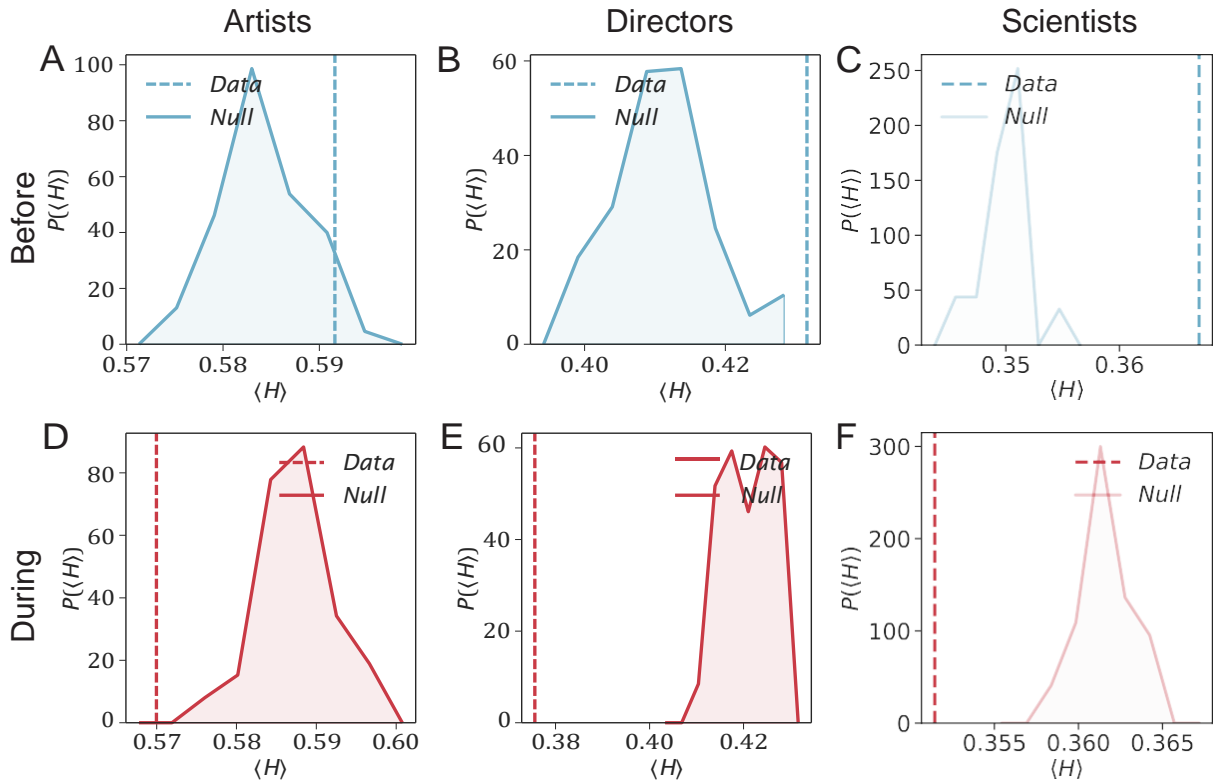

FIG. S26: The distribution of rescaled entropy $P(\langle H \rangle)$ before and during hot streak for 1000 realizations of the randomized careers measured by fixed five-year time window before and after the onset of hot streak.

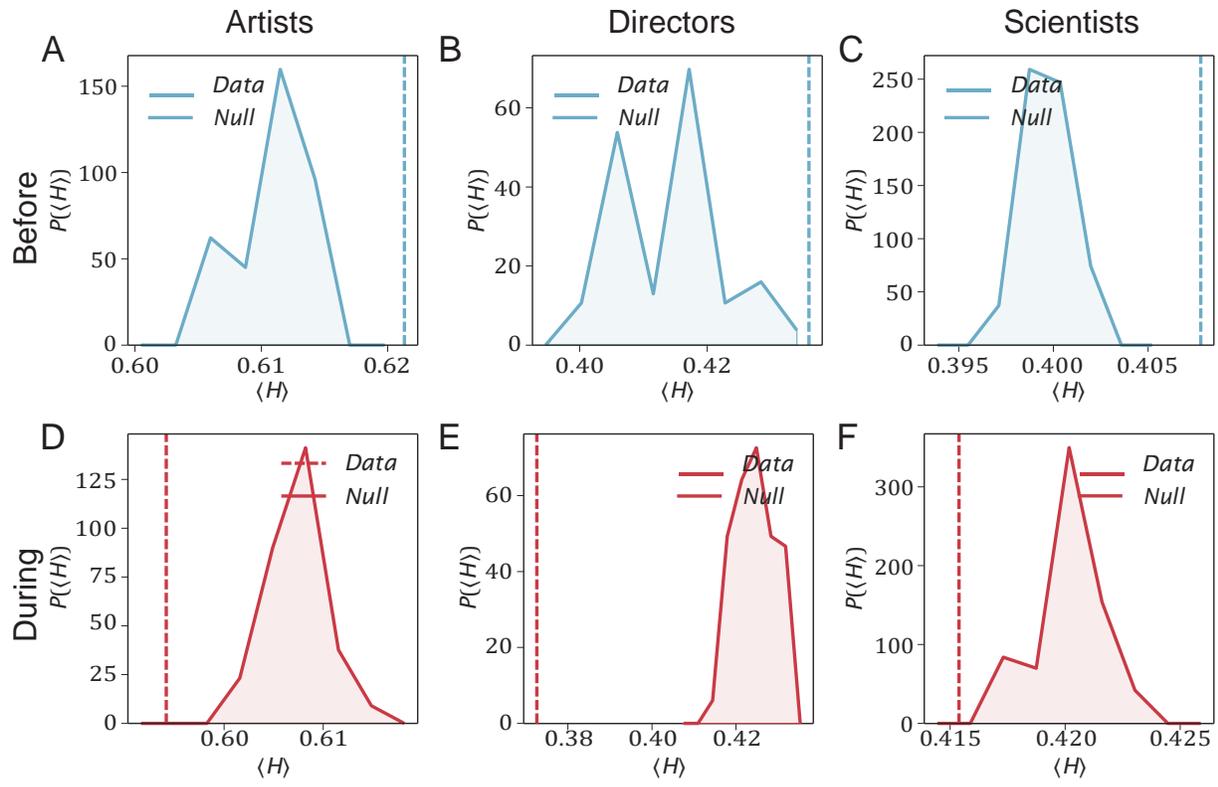

FIG. S27: The distribution of rescaled entropy $P(\langle H \rangle)$ before and during hot streak for 1000 realizations of the randomized careers measured by fixed sample size before and after the onset of hot streak (8 for artists, 5 for directors and scientists).

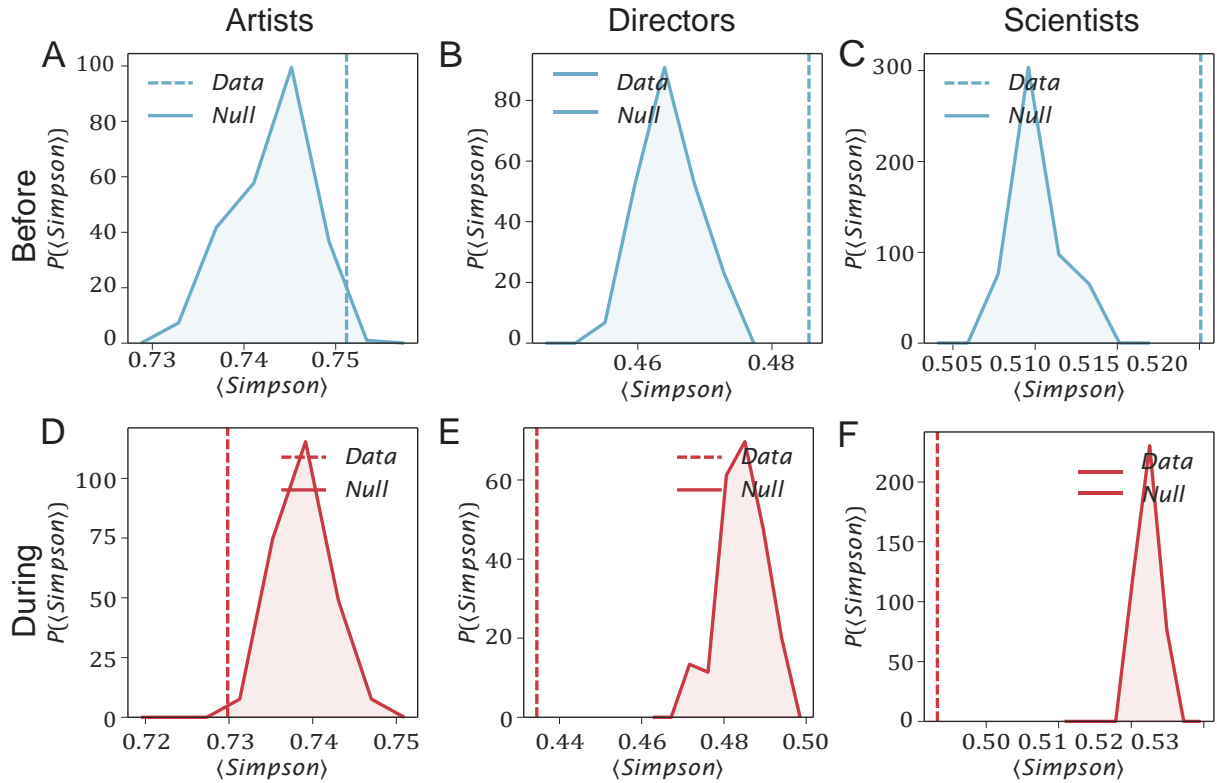

FIG. S28: The distribution of Simpsons diversity $P(\langle Simpsons \rangle)$ before and during hot streak for 1000 realizations of the randomized careers for the three domains. $\langle Simpsons \rangle$ measured from real careers (vertical line) is significantly larger than expected before hot streak (A-C) and smaller than expected during hot streak (D-F).

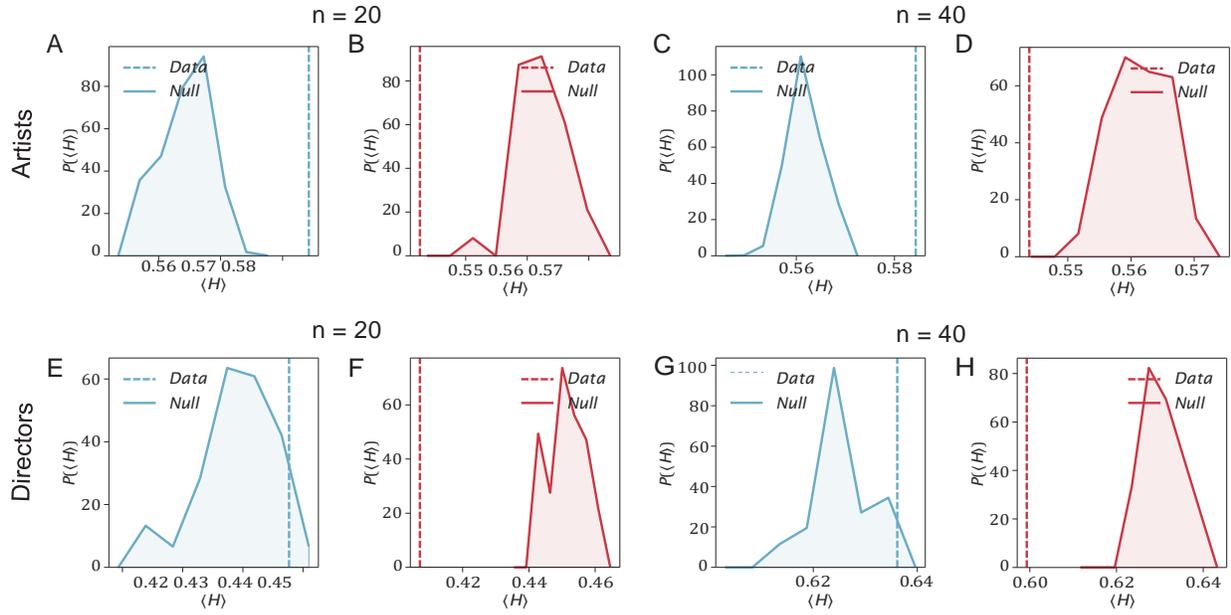

FIG. S29: The distribution of rescaled entropy $P(\langle H \rangle)$ before and during hot streak for 1000 realizations of the randomized careers using different numbers of centroids $n$ for (A-D) artists and (E-H) directors. $\langle H \rangle$ measured from real careers (vertical line) is significantly larger than expected before hot streak and smaller than expected during hot streak.

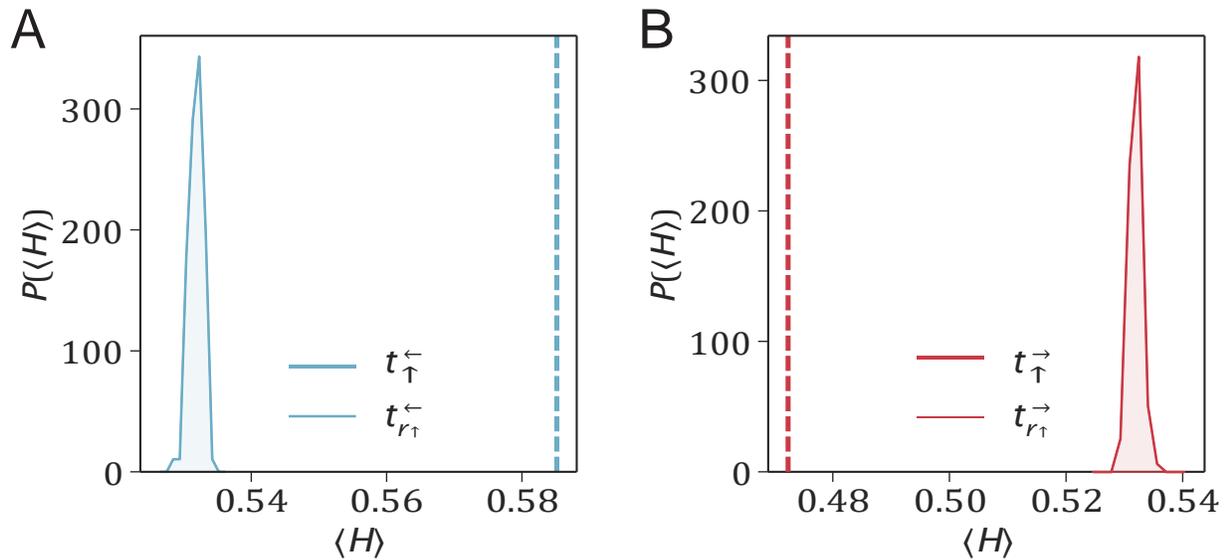

FIG. S30: The distribution of rescaled entropy $P(\langle H \rangle)$ (A) before and (B) during hot streak for 1000 realizations of the randomized scientific careers including papers without references. $\langle H \rangle$ measured from real careers (vertical line) is significantly larger than expected before hot streak and smaller than expected during hot streak.

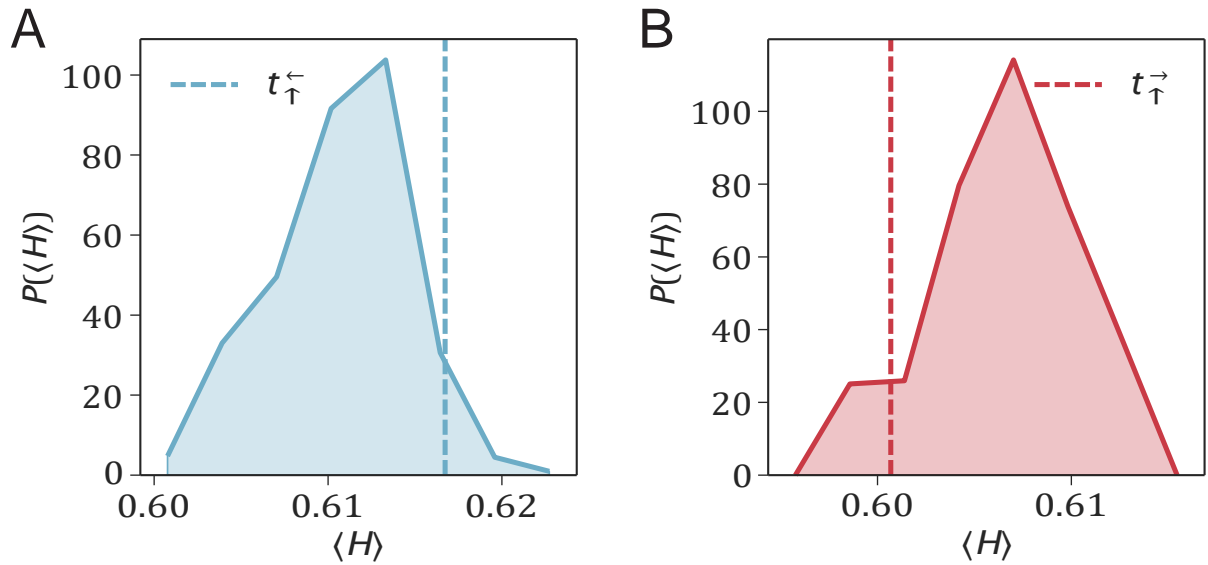

FIG. S31: The distribution of rescaled entropy $P(\langle H\rangle)$ (A) before and (B) during hot streak for 1000 realizations of the randomized artistic careers using art style labels predicted by the VGG16. $\langle H\rangle$ measured from real careers (vertical line) is significantly larger than expected before hot streak and smaller than expected during hot streak.

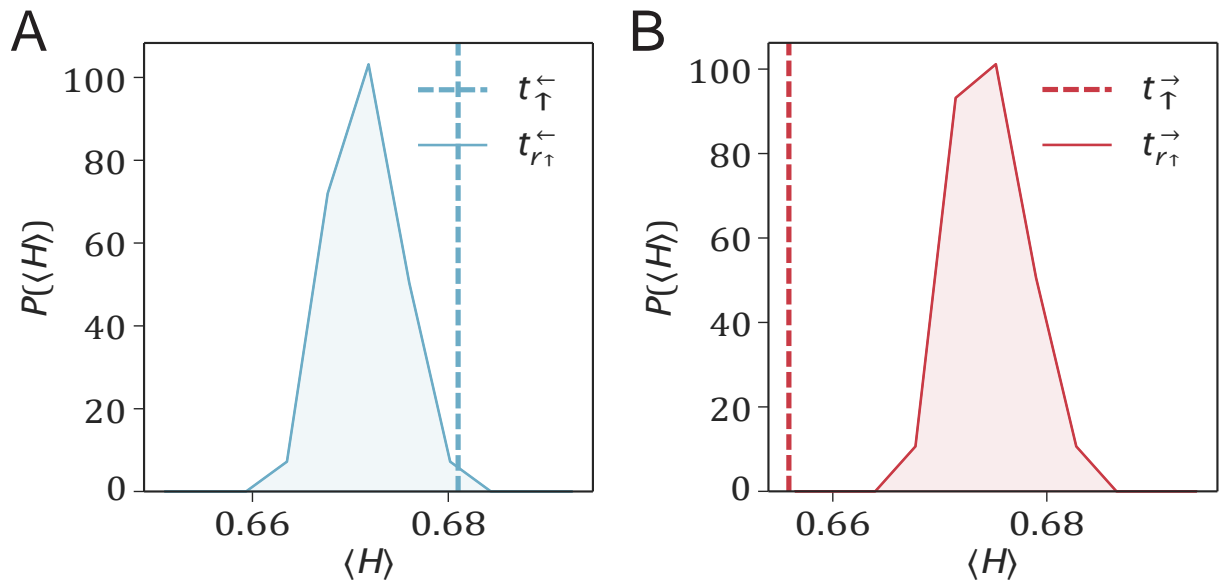

FIG. S32: The distribution of rescaled entropy $P(\langle H\rangle)$ (A) before and (B) during hot streak for 1000 realizations of the randomized directors' careers using film genre labels. $\langle H\rangle$ measured from real careers (vertical line) is significantly larger than expected before hot streak and smaller than expected during hot streak.

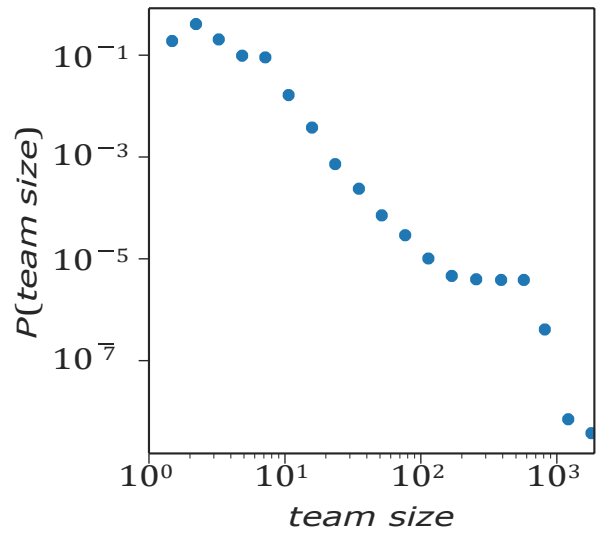

FIG. S33: Log-log plot of the team size distribution

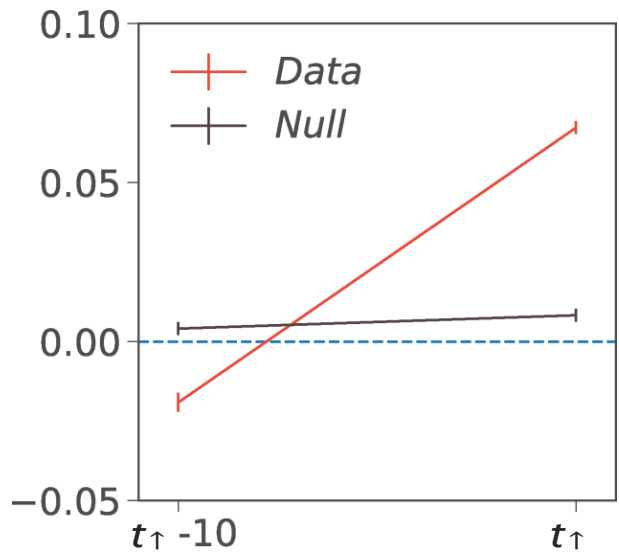

FIG. S34: Regression coefficient of team size change on the timing of hot streak from linear regressions for data and the null model.

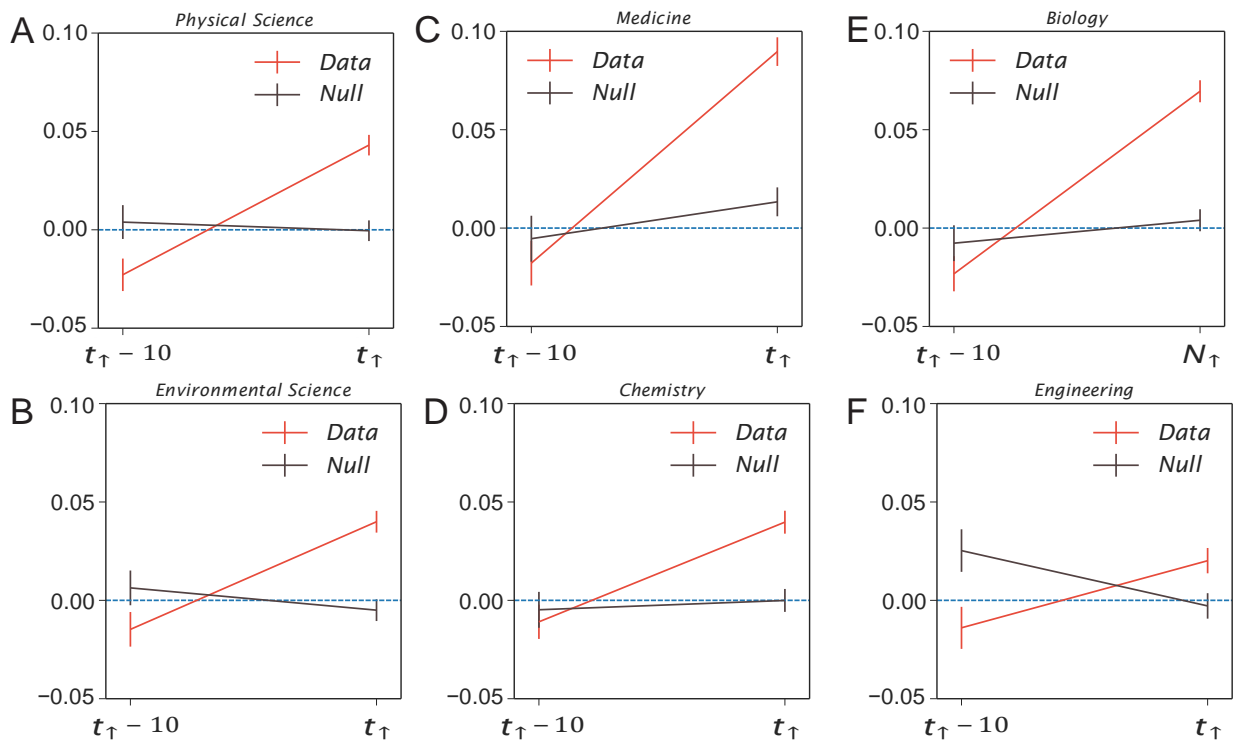

FIG. S35: Regression coefficient of team size change on the timing of hot streak from linear regressions for scientists from six different disciplines.

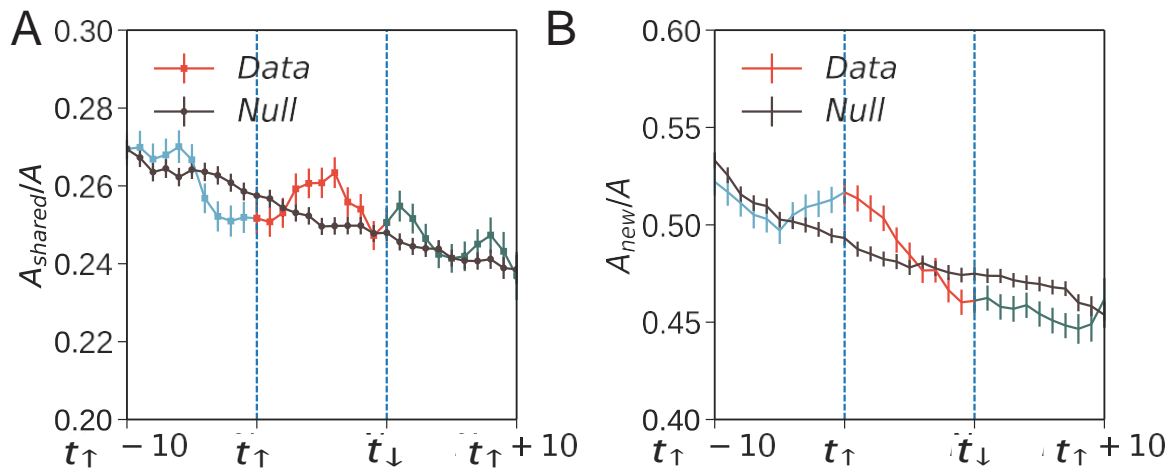

FIG. S36: (A) The turnover rate of coauthors and (B) the rate of new collaborators around hot streak.

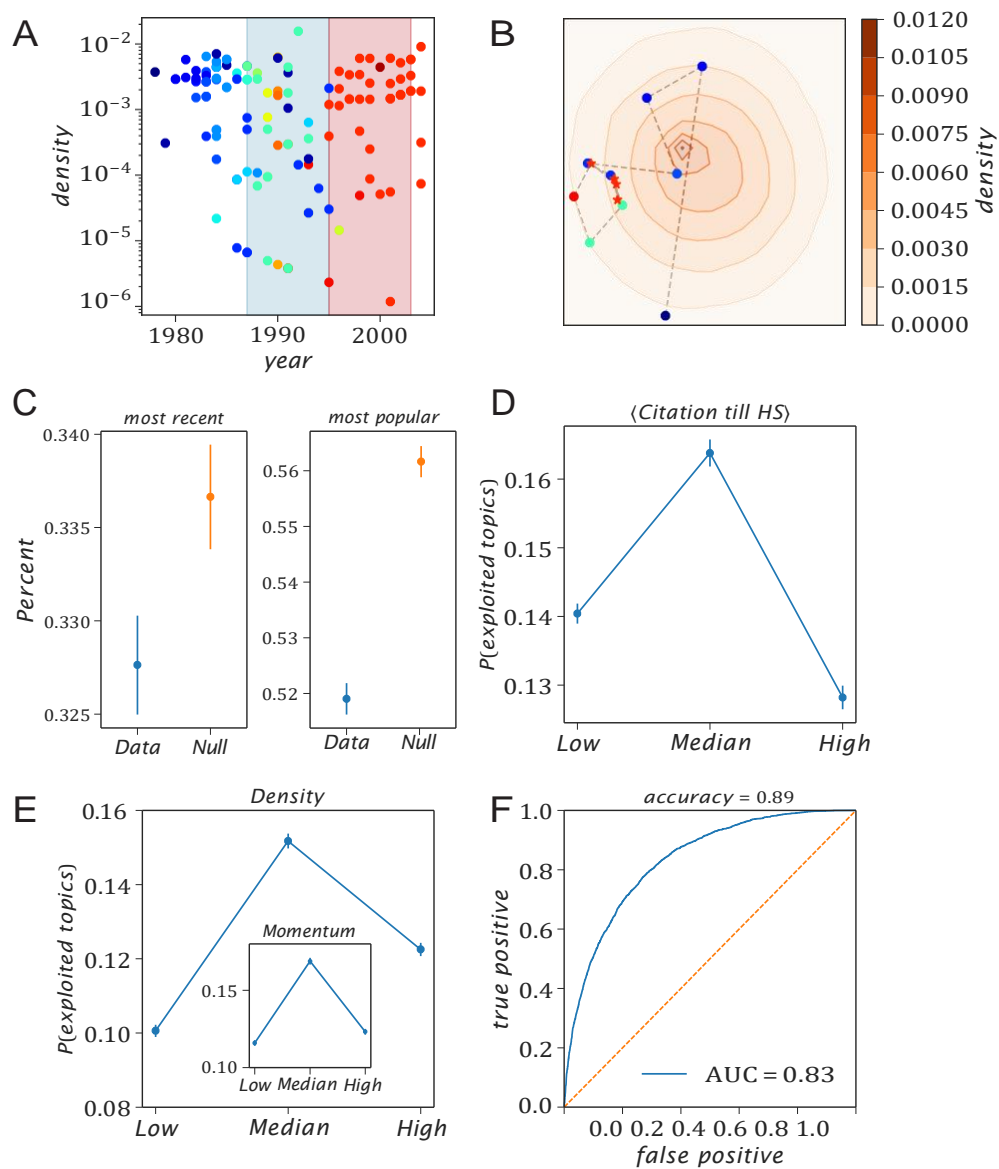

FIG. S37: (A) The dynamics of paper density within a career for a scientist in our dataset. (B) The topic trajectory in the embedding space by the same scientist in (A). Each data point denotes the average location for papers published in a two-year window. Stars denote her hot streak and dots denote the normal period. Color captures the most frequent topic in each window. (C) The probability for the topic exploited to be the (left) most recently initiated and (right) most frequently studied before a hot streak begins for real and randomized careers. (D) The probability for the topic exploited to be low-, median- and high-impact among prior topics. (E) The probability for the topic exploited to fall into low-, median- and high-density region or (Inset) momentum among prior explored topics. (F) The ROC curve for the prediction task.

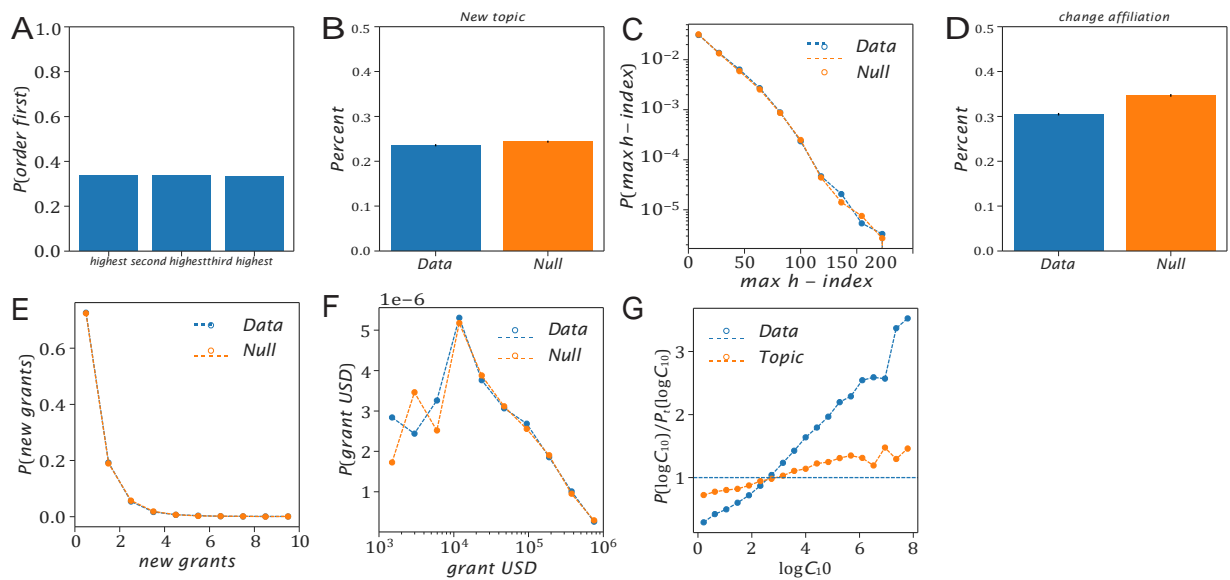

FIG. S38: (A) We measure the relative position for the top three highest impact papers during hot streak, and calculate their probability to order first among the three. We find equal probability for the three highest impact papers to appear first. (B) The probability for the hot streak to begin with a new topic for data and the null model. (C) The distribution of the highest h-index among collaborators for publications during hot streak in real and randomized careers is virtually indistinguishable. (D) The probability for an individual to work in a new institution during hot streak for data and the null model. (E) The distribution for the number of new grants at the year when a hot streak begins for data and the null model is virtually indistinguishable. (F) The distribution for new funding amount at the year when a hot streak begins for data and the null model is virtually indistinguishable. (G) The ratio between the distribution of topic impact during hot streak and that of the topics before. Compared to the ratio between the distribution of paper impact during hot streak and that of papers before, the improvement of topic impact appears negligible.